%% file: main.tex
\documentclass[conference]{IEEEtran}
\IEEEoverridecommandlockouts

\usepackage{amsthm}
\newtheorem{idea}{Idea}

\newtheorem{theorem}{Theorem}

\usepackage{enumitem}
\usepackage{xspace}
\usepackage[caption=false]{subfig}
\usepackage{multirow}
\usepackage{colortbl}
\usepackage{array}
\usepackage[linesnumbered,ruled,vlined]{algorithm2e}
\usepackage{booktabs}
\newcolumntype{C}[1]{>{\centering\arraybackslash}p{#1}}
\usepackage{soul}
\usepackage{pifont}
\usepackage{tabularx}
\usepackage{dsfont}
\usepackage{longtable}
\usepackage[hidelinks]{hyperref}
\usepackage{cite}
\usepackage{amsmath,amssymb,amsfonts}
\usepackage{graphicx}
\usepackage{textcomp}
\usepackage{xcolor}

\def\BibTeX{{\rm B\kern-.05em{\sc i\kern-.025em b}\kern-.08em
    T\kern-.1667em\lower.7ex\hbox{E}\kern-.125emX}}

\usepackage{comment}
    
\begin{document}

\title{Attributed Hypergraph Generation with Realistic Interplay Between Structure and Attributes}

\author{\IEEEauthorblockN{Jaewan Chun\IEEEauthorrefmark{1}, Seokbum Yoon\IEEEauthorrefmark{1}, Minyoung Choe, Geon Lee, Kijung Shin}
\IEEEauthorblockA{\textit{Kim Jaechul Graduate School of AI, KAIST, Seoul, Republic of Korea} \\}
\IEEEauthorblockA{\{jjwpalace, jing9044, minyoung.choe, geonlee0325, kijungs\}@kaist.ac.kr \\}
}

\maketitle
\begingroup\renewcommand\thefootnote{\IEEEauthorrefmark{1}}
\footnotetext{Equal Contributions}
\endgroup

\maketitle

\input{dfn.tex}

\input{000abstract}


\section{Introduction}
\label{sec:intro}
\input{010intro.tex}

\section{Related Work}
\label{sec:related}
\input{020related.tex}

\section{Preliminaries}
\label{sec:prelim}
\input{030prelim.tex}

\section{Proposed Generation Method: \method}
\label{sec:method}
\input{040method.tex}

\section{Proposed Fitting Method: \fitwattr}
\label{sec:fitting}
\input{050fitting}
\section{Experiments}
\label{sec:experiments}
\input{060exp.tex}

\section{Conclusions}
\label{sec:conclusion}
\input{070conclusion.tex}

\section*{Acknowledgements}
{\small This work was supported by Institute of Information \& Communications Technology Planning \& Evaluation (IITP) grant funded by the Korea government (MSIT)  (No. 2022-0-00871 / RS-2022-II220871, Development of AI Autonomy and Knowledge Enhancement for AI Agent Collaboration, 50\%)
(No. RS-2024-00457882, AI Research Hub Project, 40\%)
(RS-2019-II190075, Artificial Intelligence Graduate School Program (KAIST), 10\%).}

\bibliographystyle{IEEEtran}
\bibliography{ref}



\end{document}

%% file: dfn.tex
\newcommand\red[1]{\textcolor{black}{#1}}
\newcommand\blue[1]{\textcolor{black}{#1}}
\newcommand\green[1]{\textcolor{black}{#1}}
\newcommand\brown[1]{\textcolor{black}{#1}}
\newcommand\purple[1]{\textcolor{black}{#1}}

\newcommand\violet[1]{\textcolor{violet}{#1}}

\definecolor{algblue}{RGB}{0, 0, 255}
\newcommand\algblue[1]{\textcolor{algblue}{#1}}
\newcommand{\algblueline}[1]{\textcolor{algblue}{\hfill\footnotesize$\blacktriangleright$~#1}}

\newcommand{\smallsection}[1]{{\noindent {\bf{\underline{\smash{#1}}}}}}

\newcommand{\method}{\textsc{NoAH}\xspace}
\newcommand{\fitwattr}{\textsc{NoAHFit}\xspace}
\newcommand{\hypercl}{\textsc{HyperCL}\xspace}
\newcommand{\hyperpa}{\textsc{HyperPA}\xspace}
\newcommand{\hyperff}{\textsc{HyperFF}\xspace}
\newcommand{\hyperlap}{\textsc{HyperLAP}\xspace}
\newcommand{\hyperdk}{hyper d\textsc{K}-series\xspace}
\newcommand{\thera}{\textsc{THera}\xspace}
\newcommand{\hycosbm}{\textsc{HyCoSBM}\xspace}
\newcommand{\hyrec}{\textsc{HyRec}\xspace}
\newcommand{\methodm}{\textsc{EDGE}$_\mathrm{M}$\xspace}
\newcommand{\methodd}{\textsc{EDGE}$_\mathrm{D}$\xspace}
\newcommand{\vect}[1]{\boldsymbol{#1}}
\newcommand{\mat}[1]{\boldsymbol{\rm #1}}

\definecolor{rankone}{RGB}{0,176,240}
\definecolor{ranktwo}{RGB}{0,255,0}
\definecolor{rankthree}{RGB}{255,255,0}
\definecolor{textyellow}{rgb}{1, 0.86275, 0.23529}

\definecolor{real}{RGB}{200,200,200}   
\definecolor{method}{RGB}{214,39,40}

\definecolor{verylightgray}{gray}{0.9}

\definecolor{orange}{RGB}{255, 170, 51}
\colorlet{transorange}{orange!50}
\newcommand{\hlorange}[1]{{\sethlcolor{transorange}\hl{#1}}}

\definecolor{blue}{RGB}{58, 102, 177}
\colorlet{transblue}{blue!50}
\newcommand{\hlblue}[1]{{\sethlcolor{transblue}\hl{#1}}}

\definecolor{tabblue}{RGB}{31, 119, 180}
\colorlet{tabblue}{tabblue!50}
\newcommand{\tabblue}[1]{{\sethlcolor{tabblue}\hl{#1}}}

\definecolor{tabred}{RGB}{214, 39, 40}
\colorlet{tabred}{tabred!50}
\newcommand{\tabred}[1]{{\sethlcolor{tabred}\hl{#1}}}

\definecolor{tabgray}{RGB}{127, 127, 127}
\colorlet{tabgray}{tabgray!50}
\newcommand{\tabgray}[1]{{\sethlcolor{tabgray}\hl{#1}}}

\setlength{\textfloatsep}{0.12cm}
\setlength{\dbltextfloatsep}{0.12cm}
\setlength{\abovecaptionskip}{0.12cm}
\setlength{\skip\footins}{0.12cm}

\newcolumntype{Y}[1]{>{\centering\arraybackslash}p{#1}}

\let\oldnl\nl
\newcommand{\nonl}{\renewcommand{\nl}{\let\nl\oldnl}}

%% file: 000abstract.tex
\begin{abstract}
In many real-world scenarios, interactions happen in a group-wise manner with multiple entities, and therefore, hypergraphs are a suitable tool to accurately represent such interactions.
Hyperedges in real-world hypergraphs are not composed of randomly selected nodes but are instead formed through \blue{structured} processes. 
Consequently, various hypergraph generative models have been proposed to \blue{explore} fundamental mechanisms underlying hyperedge formation.
However, most existing hypergraph generative models do not account for node attributes, which can play a significant role in hyperedge formation. 
As a result, these models fail to \blue{reflect} the interactions between structure and node attributes.

To address the issue above, we propose \method, a stochastic hypergraph generative model for attributed hypergraphs.
\method utilizes the core–fringe node hierarchy to model hyperedge formation as a series of node attachments and determines attachment probabilities based on node attributes.
\blue{We further introduce \fitwattr, a parameter learning procedure that allows \method to replicate a given real-world hypergraph.}
Through experiments on nine datasets across four different domains, we show that \method with \fitwattr \blue{more accurately reproduces the structure–attribute interplay observed in the real-world hypergraphs than eight baseline hypergraph generative models, in terms of six metrics.}
\end{abstract}

%% file: 010intro.tex
Many real-world interactions occur in groups, such as co-authorship among researchers, group discussions on online Q\&A sites, and co-purchasing of items. 
Hypergraphs, which consist of 
\blue{hyperedges,} naturally and effectively represent group interactions involving an arbitrary number of individuals or entities. 
\blue{Especially,
hypergraph modeling} has shown effectiveness in a variety of applications, including clustering\cite{kumar2020hypergraph, hayashi2020hypergraph}, classification~\cite{yu2012adaptive}, and anomaly detection~\cite{silva2008hypergraph, chun2024random}.

Hyperedges in real-world hypergraphs are not composed of random nodes but are generally formed in a more \blue{systematic} manner.
\red{For instance, real-world hypergraphs often exhibit high-degree nodes~\cite{do2020structural}, densely overlapping hyperedges~\cite{lee2021hyperedges}, and high transitivity~\cite{kim2023transitive}.}

Building upon these findings, a number of hypergraph generative models have been proposed, incorporating hyperedge generation mechanisms \blue{that lead to realistic hypergraph structures.}
These hypergraph generation models allow a better understanding of real-world hypergraphs, and they are also employed in \blue{various data-mining} applications, including community detection~\cite{ghoshdastidar2017consistency, ruggeri2023community, badalyan2024structure}, and hyperedge prediction~\cite{contisciani2022inference}.

Despite the success of hypergraph generative models, they mostly overlook interplays between hypergraph structure and node attributes.
Node attributes are commonly associated with real-world data. 
For example, in co-authorship hypergraphs, 
where nodes represent 
authors and hyperedges represent co-authored publications, node attributes, such as affiliation and field of study, offer valuable information.
\red{Especially, as exemplified by homophily~\cite{mcpherson2001birds, lee2024villain}, such node attributes can influence the formation of collaborations (i.e., hyperedges).}

Thus, in this paper, we propose \method (\underline{\smash{\textbf{No}}}de \underline{\smash{\textbf{A}}}ttribute based \underline{\smash{\textbf{H}}}ypergraph generator), a novel hypergraph generative model based on node attributes. 
Since a hyperedge can involve an arbitrary number of nodes, the number of hyperedge candidates increases exponentially with the number of nodes.
As a result, generating a hypergraph by considering the formation probabilities of all candidates is computationally intractable.
To address this challenge, \method models the formation of each hyperedge as a series of attachments of nodes to its seed node(s).
The attachment probabilities are determined by node attributes.
Specifically, we assign the degree of affinity based on the values of each node attribute, and we obtain the final attachment probability as the product of the affinity scores across all attributes.
\blue{In addition, \method incorporates a core–fringe node hierarchy into the process to enhance realism.}

We also introduce \fitwattr, an algorithm designed to fit the parameters of \method to a given hypergraph.
The hyperedge formation probabilities in \method are expressed through a parameterized formulation, and \fitwattr updates the parameters to maximize the probabilities,
capturing the structure-attribute interplay in the given hypergraph.

\begin{figure*}[t]
    \vspace{-3mm}
    \centering
    \includegraphics[width=2\columnwidth]{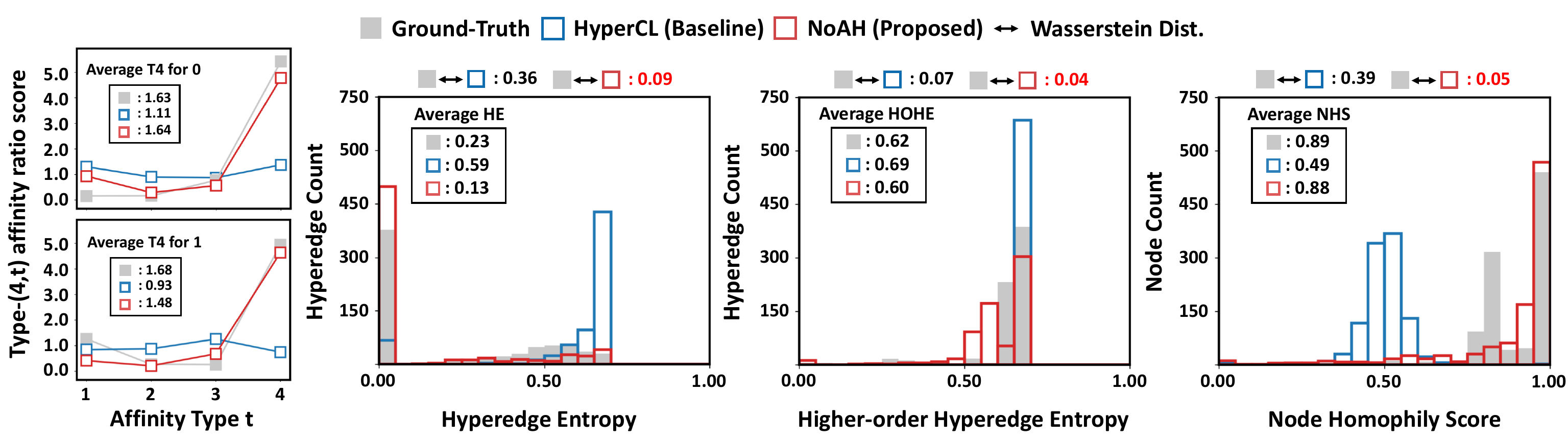}
    \vspace{-1mm}
    \caption{
    \red{Structure–attribute interplay with respect to the first node attribute in the Amazon Music dataset.
    Our proposed generative model, \method, fitted by its parameter learning algorithm \fitwattr, effectively captures the interplay between structure and node attributes, outperforming the baseline model (HyperCL).
    Refer to Section~\ref{sec:prelim:measures} for the definitions of the measures and to Section~\ref{sec:exp:case} for the interpretation of the results.
    } 
    }
    \label{fig:intro_example}
\end{figure*}

In our experiments, we extensively evaluate hypergraph generative models using six measures on their ability to capture structure–attribute interplay in nine real-world hypergraphs.
As exemplified in Figure~\ref{fig:intro_example},  \method, fitted by \fitwattr, outperforms all eight existing hypergraph \red{generative models} in the overall assessment across the measures. 

Our contributions are summarized as follows:
\begin{itemize}[leftmargin=*]
\item \textbf{Model:} We propose \method, a stochastic generative model for attributed hypergraphs that produces a realistic interplay between structure and attributes.
\item \textbf{Fitting Algorithm:} We develop \fitwattr, a parameter fitting algorithm for \method that 
captures the relationship between structure and node attributes in a given hypergraph to maximize the formation probabilities of its hyperedges.
\item \textbf{Experiments:} \red{We empirically show that \method better reproduces the structure–attribute interplay in real-world hypergraphs than eight baseline hypergraph generative models.}
\end{itemize}
For \textbf{reproducibility}, we make the source code \red{and data} publicly available at \url{https://github.com/jaewan01/NoAH}.

%% file: 020related.tex
In this section, we review prior work on the generation of graphs and hypergraphs.

\subsection{(Hyper)graph Generative Models}
(Hyper)graph generative models are designed to generate structures resembling real-world hypergraphs~\cite{chakrabarti2006graph, lee2024survey}, often to uncover potential mechanisms behind such structures.
To this end, for example, several models exploit node hierarchies~\cite{jia2019random, kim2023transitive,papachristou2022core, ko2022growth}, which are commonly found in the real world~\cite{seidman1983network, borgatti2000models, bu2023hypercore, tudisco2023core}.
In our model, \method, we capture such hierarchies through a simple yet effective core–fringe structure.

Recently, various deep learning based (hyper)graph generative models have been introduced, 
such as variational autoencoders~\cite{simonovsky2018graphvae}, generative adversarial networks~\cite{bojchevski2018netgan,de2018molgan}, and diffusion or hierarchical models~\cite{liu2023diffusion}.
However, these models mostly require a large collection of graph instances for training, 
while the aforementioned models operate on a single instance.
Moreover, deep learning models often struggle to provide insights into the generative mechanisms of real-world (hyper)graphs due to their intrinsic black-box nature.

\subsection{Attributed-aware (Hyper)graph Generative Models}
Node attributes can play a crucial role in graph generation, as demonstrated by homophily, a prevalent property of real-world (hyper)graphs where similar nodes tend to connect~\cite{mcpherson2001birds}.
Accordingly, several models incorporate node attributes, including the exponential random graph model~\cite{robins2007exp}, the stochastic block model~\cite{airoldi2008mixed}, and the latent space model~\cite{wang2023joint}.
Among them, the Multiplicative Attribute Graph (\textsc{MAG}) model\cite{kim2012multiplicative} is distinctive in that node attributes directly govern edge formation via a multiplicative probability function.

However, 
\blue{very few} hypergraph generative models incorporate node attributes. 
\blue{One exception is} a stochastic block model variant specifically designed for community detection~\cite{badalyan2024structure}. 
In this model, node attributes are conditionally independent of the hypergraph structure given the latent community structure, 
meaning attributes do not directly drive hyperedge formation.
In contrast, in our model, \method, hyperedges are formed explicitly based on attribute relationships among nodes.

%% file: 030prelim.tex
In this section, we introduce (1) key notations, (2) six measures for evaluating the interplay between structure and attributes, (3) MAG~\cite{kim2012multiplicative} as a preliminary model, and (4) UMHS \cite{amburg2021planted} as a core-node identification method for hypergraphs.

\setlength{\tabcolsep}{4pt} 
\begin{table}[t!]
    \centering
    \caption{\centering Frequently-used symbols.}
    \label{tab:symbols}
    \begin{tabular}{ m{11em} m{18em}}
        \toprule
        Notation & Definition \\
        \midrule
        $\mathcal{H}=(\mathcal{V}, \mathcal{E}, \mat{X})$ & hypergraph with nodes $\mathcal{V}$, hyperedges $\mathcal{E}$ \\
        $\mat{X} \in {\{0,1\}}^{\vert \mathcal{V} \vert \times k}$ & node attribute matrix (binary)\\
        \midrule
        $\mathcal{C}$ & set of core nodes\\
        $\mathcal{F}$ & set of fringe nodes\\
        $\Theta_{\mathcal{C}}=\{\mat{\theta}_{\mathcal{C}_1}, \dots, \mat{\theta}_{\mathcal{C}_k} \}$ & set of  \purple{core group} affinity matrices\\
        $\Theta_{\mathcal{F}}=\{\mat{\theta}_{\mathcal{F}_1}, \dots, \mat{\theta}_{\mathcal{F}_k} \}$ & set of  \purple{fringe attachment} affinity matrices \\
        \bottomrule
    \end{tabular}
\end{table}

\subsection{Notations}\label{sec:prelim:notations}
First, we discuss the notations used in this paper.
Refer to Table~\ref{tab:symbols} for the frequently-used notations.

\smallsection{Attributed Hypergraphs.}
An \textit{attributed hypergraph} $\mathcal{H}=(\mathcal{V}, \mathcal{E}, \mat{X})$ consists of a set of nodes $\mathcal{V} = \{v_1,\cdots,v_{\vert \mathcal{V} \vert} \}$, a set of hyperedges $\mathcal{E} = \{e_1,\cdots,e_{\vert \mathcal{E} \vert} \}$, and a node attribute matrix $\mat{X} \in \mathbb{R}^{\vert \mathcal{V} \vert \times k}$. 
Each hyperedge $e\in\mathcal{E}$ is a non-empty subset of nodes, i.e., $e \subseteq \mathcal{V}, \vert e \vert \geq 1$. 
The $i$-th row of $\mat{X}$, denoted as $\mathbf{x}_i=\mat{X}_{i,:}\in \mathbb{R}^k$, represents the attribute vector of node $v_i\in \mathcal{V}$.
We use $\mathbf{x}_i^{(l)}\in \mathbb{R}$ to denote the $l$-th attribute value of node $v_i$.
In this work, we assume binary node attributes, i.e., $\mat{X}\in \{0,1\}^{|\mathcal{V}|\times k}$, 
which simplifies both model design and implementation while remaining valid for many real-world datasets.
Moreover, categorical and continuous attributes can also be converted into binary ones via one-hot encoding and thresholding, respectively, as in our experiments.

\subsection{Measures for Structure-Attribute Interplay}\label{sec:prelim:measures}
We introduce several measures for evaluating the interplay between structural patterns and node attributes,  
categorized into:
(1) \textbf{hyperedge-level measures} (hyperedge entropy and affinity ratio score), which capture how node attributes are distributed within hyperedges, and a 
(2) \textbf{node-level measure} (node homophily score), which measures the tendency of nodes to form hyperedges with others of similar attributes.

\smallsection{Type-$\boldsymbol{s}$ Affinity Ratio Scores.} Veldt et al.~\cite{veldt2023combinatorial} introduced a mathematical framework to quantify the \red{significance} of a label on group interactions of a fixed size $s$. First of all, for each affinity type $t \in \{1, 2, \dots, s\}$, they defined the type-($s$, $t$) \textit{affinity score} for label $Y$ as follows:
\begin{equation}
    h_{s,t}(Y) := \frac{\sum_{v\in \mathcal{V}_Y}d_{s,t}(v)}{\sum_{v\in \mathcal{V}_Y}d(v)}, \label{eq:affinity_score} \vspace{-1mm}
\end{equation}
where $\mathcal{V}_Y$ is the set of nodes with label $Y$, $d(v)$ the degree of node $v$, and $d_{s,t}(v)$ the number of size $s$ hyperedges containing $v$ and $t-1$ additional nodes with $v$'s label \violet{($Y$)}. 
It \red{quantifies} how frequently nodes with label $Y$ appear in size-$s$ hyperedges with exactly $t$ such nodes.
\red{The type-($s$, $t$) \textit{baseline score} $b_{s,t}(Y)$ for label $Y$ is the probability that a node with label $Y$ joins a size-$s$ hyperedge containing exactly $t$ nodes with label $Y$, if $s-1$ other nodes are selected uniformly at random, i.e.,} 
\vspace{-1mm}
\begin{equation}
    b_{s,t}(Y) := \frac{\binom{\vert \mathcal{V}_Y \vert-1}{t-1}\binom{\vert \mathcal{V} \vert-\vert \mathcal{V}_Y \vert}{s-t}}{\binom{\vert \mathcal{V} \vert-1}{s-1}}. \label{eq:baseline_score} \vspace{-1mm}
\end{equation}
\red{We examine} the type-($s$, $t$) \textit{affinity ratio score} for label $Y$, \red{which is} the ratio of the affinity score to the baseline score, i.e., $\frac{h_{s,t}(Y)}{b_{s,t}(Y)}$, to evaluate the \red{significance} of the combination of $t$ nodes labeled $Y$ in a fixed hyperedge size $s$.

Since we assume binary attributes, we consider the type-($s$, $t$) affinity ratio score for values $0$ and $1$ of each attribute.
In this work, we focus on $s \in \{2, 3, 4\}$, as \red{small hyperedges} are common in many real-world hypergraphs, and to allow reliable statistics. 
For simplicity, we refer to the type-($s$, $1$) to type-($s$, $s$) affinity ratio scores as the type-$s$ affinity ratio scores. 

\smallsection{(Higher-order) Hyperedge Entropy.} 
Lee et al.~\cite{lee2024villain} observed that real-world hyperedges exhibit label homogeneity, containing nodes with similar labels more often than their randomized counterparts.
This effect persists even after multiple steps of propagation of labels \red{to} incident nodes and hyperedges.
To measure the label homogeneity, they utilized \textit{hyperedge entropy},
defined as the entropy of the labels of nodes incident to each hyperedge.
They also introduced \textit{higher-order hyperedge entropy}, measuring the entropy after label propagation.

In this work, we consider the distribution of hyperedge entropy and higher-order hyperedge entropy of each \red{node attribute} to measure the \red{dominance} of attributes on hyperedge formation. A distribution skewed toward 0 reflects that hyperedges mostly comprise nodes with identical attribute values.

\smallsection{Node Homophily Score.} We propose a \textit{node homophily score} to quantify the homogeneity of nodes in a hypergraph. Node homophily score of node $v$ at $l$-th attribute is defined as:
\begin{equation}
    n_v[l] := \frac{\sum_{e\in\mathcal{E}_v}\vert\{u\in e, u\neq v\;\vert\;\mat{X}[u,l]=\mat{X}[v,l]\}\vert}{\sum_{e\in\mathcal{E}_v}(\vert e \vert-1)}, \label{eq:node_homophily_score} \vspace{-1mm}
\end{equation}
where $n_v[l]$ is a node homophily score of node $v$ for \red{the $l$-th attribute}, $\mathcal{E}_v \subseteq \mathcal{E}$ is a set of hyperedges containing $v$, and $l\in\{1, 2, \dots, k\}$. 
This measure indicates the \purple{ratio} of incident nodes that share the same $l$-th attribute value with node $v$, 
with a high value reflecting a greater tendency for $v$ to form hyperedges with nodes having the same attribute value.
We consider the distribution of node homophily score \red{for each attribute} to capture node-level homogeneity in a hypergraph. 

Overall, the above measures offer complementary perspectives on structure-attribute interplay:  (1) the \textbf{type-$s$ affinity ratio score} \red{quantifies} the fine-grained patterns in hyperedge-attribute distributions, (2) the \textbf{(higher-order) hyperedge entropy} \red{quantifies} the coarse-grained patterns in hyperedge-attribute distributions, and (3) the \textbf{node homophily score} \red{quantifies} the node-level patterns of attribute distributions.

\subsection{MAG: Multiplicative Attribute Graph Model}\label{sec:prelim:mag}
Kim and Leskovec~\cite{kim2012multiplicative} proposed the Multiplicative Attribute Graph (\textsc{MAG}) model to capture how node attributes affect edge formation in pairwise graphs.  
Given a set of nodes $\mathcal{V}$ and an associated node attribute matrix $\mat{X} \in {\{0,1\}}^{\vert \mathcal{V} \vert \times k}$, MAG estimates the probability of edges based on node attributes.~\footnote{While MAG can consider a general categorical attribute, in most cases simplified MAG model with binary attribute is utilized.}
Specifically, MAG defines a set of attribute affinity matrices $\Theta=\{\mat{\theta}_1, \dots, \mat{\theta}_k \}$, where each 
$\mat{\theta}_l \in \mathbb{R}^{2\times 2}$ 
captures the affinity between values of the $l$-th attribute, with $\mat{\theta}_l[x_1,x_2] \in [0,1]$ denoting the affinity between binary attribute values $x_1,x_2\in \{0,1\}$.
For the probability $P(v_i,v_j)$ of an edge between nodes $v_i$ and $v_j$, MAG multiplies the affinities across all $k$ attributes as follows:
\begin{equation}
    P(v_i,v_j) = \prod\nolimits_{l=1}^k \mat{\theta}_{l}[\mathbf{x}_i^{(l)}, \mathbf{x}_j^{(l)}].
    \vspace{-1mm}
    \label{eq:mag}
\end{equation}
Inspired by MAG, 
we develop a model for group interactions, which require more complex modeling of how node attributes collectively influence group formation. 
\red{Moreover, compared to MAG,
our model additionally captures commonly-observed structural node hierarchies} by distinguishing between core and fringe node roles in group formation.

\begin{figure*}[!t] 
    \vspace{-5mm}
    \centering
    \includegraphics[width=2\columnwidth]{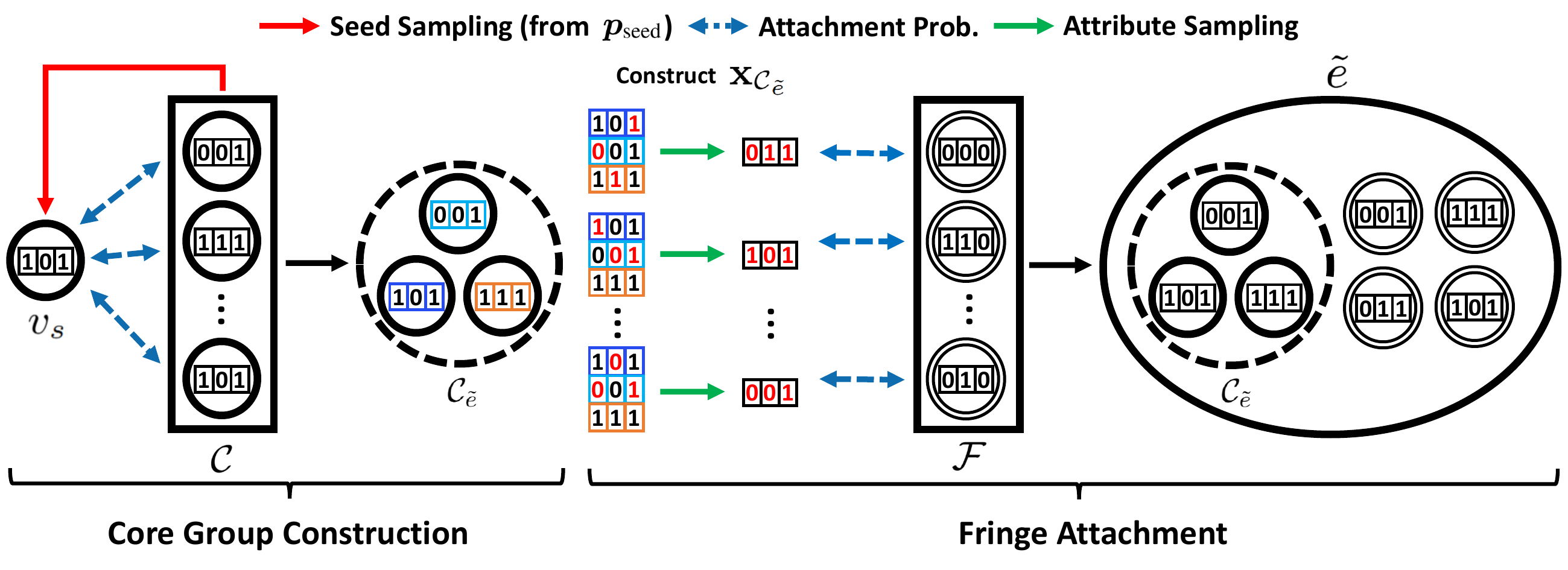}
    \vspace{-3mm}
    \caption{Hyperedge generation process of \method consists of two steps: (1) Core group construction: sample a seed core node $v_s$ according to $\vect{p}_\text{seed}$ and attach additional core nodes to $v_s$ to form a core group \blue{$\mathcal{C}_{\tilde{e}}$}, and (2) Fringe attachment: attach fringe nodes to core group \blue{$\mathcal{C}_{\tilde{e}}$} based on mixed attribute vector \blue{ $\vect{x}_{C_{\Tilde{e}}}$} obtained by attribute-wise sampling from \blue{$\mat{X}_{\mathcal{C}_{\tilde{e}}}$}.
    \red{The attached core and fringe nodes, together with the seed node, form a hyperedge.}
    }
    \label{fig:method}
\end{figure*}

\subsection{UMHS: Union of Minimal Hitting Sets}\label{sec:prelim:umhs}
Amburg et al.~\cite{amburg2021planted} defined a \textit{hitting set} of a hypergraph as a set of nodes that intersects every hyperedge, i.e., a set $S$ satisfying $\forall e \in \mathcal{E}$, $\exists v \in S$ s.t. $v\in e$.
Building on this concept, they proposed the union of minimal hitting sets (UMHS) algorithm to identify core nodes in a hypergraph.
The algorithm consists of three steps:
(1) initialize $S=\{\}$, generate a random permutation of the hyperedges, and traverse them, adding all nodes in a hyperedge $e$ to $S$ whenever $S \cap e = \emptyset$, (2) generate a random permutation of $S$ and traverse them, removing any node $v$ if $S \backslash \{ v \}$ remains a hitting set, and (3) repeat the above two steps with different permutations and take the union of the resulting hitting sets.

%% file: 040method.tex
In this section, we introduce \method, a novel hypergraph generative model based on node attributes.

\subsection{Ideas behind \method.} \label{sec:method:idea}
In real-world hypergraphs, node attributes play a crucial role in hyperedge formation. 
\red{For example, in some hypergraphs, hyperedges among nodes with similar attributes are common (homophily) \cite{lee2024villain}, while in others, hyperedges among nodes with dissimilar attributes (heterophily) are common \cite{li2025hypergraph}. 
\method captures such attribute-structure interplay to generate more realistic hypergraphs.}

\begin{idea} \label{idea:attr}
    We model hyperedge formation using node attributes to reflect their interplay with structure.
\end{idea}

Moreover, many real-world hypergraphs exhibit hierarchical structures~\cite{bu2023hypercore, tudisco2023core}, where certain nodes consistently play central roles in group interactions, while others interact more peripherally. This hierarchy has been incorporated into several generative models~\cite{ko2022growth, papachristou2022core, kim2023transitive}. 
In \method, we \red{introduce a surprisingly} simple yet effective approach to \red{reflect} hierarchy by dividing nodes into two groups with distinct structural roles. 
\begin{idea} \label{idea:corefringe}
    We divide nodes into core and fringe nodes to capture the hierarchical structure commonly observed in real-world hypergraphs.
    Core nodes play a more central role in hyperedge formation, while fringe nodes participate more peripherally.
\end{idea}

\red{Since the number of hyperedge candidates increases exponentially with the number of nodes, generating a hypergraph by considering the formation probabilities of all candidates is computationally intractable.
To address this challenge, \method incrementally constructs each hyperedge through attachment.}
\begin{idea} \label{idea:seq}
    \red{We model the formation of each hyperedge as a series of attachments of nodes to its seed node(s).}
\end{idea}



\subsection{Model Details of  \method.} \label{sec:method:detail}
\red{Based on Idea~\ref{idea:seq},} \method generates a hypergraph by stochastically sampling nodes to attach to each hyperedge, as \red{outlined in Algorithm~\ref{alg:method} and illustrated in Figure~\ref{fig:method}.}
In line with Idea~\ref{idea:attr}, \red{nodes are sampled} based on attribute relationships, using the node attributes $\mathbf{X}$.
To incorporate Idea~\ref{idea:corefringe}, the node set $\mathcal{V}$ is partitioned into two disjoint subsets: the core node set $\mathcal{C}\subseteq \mathcal{V}$ and the fringe node set $\mathcal{F}\subseteq \mathcal{V}$, s.t. $\mathcal{C} \cap \mathcal{F} = \varnothing$ and $\mathcal{C} \cup \mathcal{F} = \mathcal{V}$, and \red{the two groups play distinct roles in hyperedge formation.}
\red{Combining three ideas,}  \method generates each hyperedge in a two-step process.
\textit{First}, a subset of core nodes from $\mathcal{C}$ form a core group based on attribute affinities.
\textit{Second}, fringe nodes from $\mathcal{F}$ attach to the core group according to attribute affinity with the group, completing the hyperedge.

\begin{itemize}[leftmargin=*]
    \item \textbf{Step 1. Core Group Construction (lines \ref{alg:method:cgstart} -- \ref{alg:method:cgend}):}
    To initiate hyperedge construction, \method begins by forming a subset of core nodes $\mathcal{C}_{\tilde{e}}\subseteq \mathcal{C}$, which plays a role as the \textit{structural nucleus} of the hyperedge.
    Specifically, it first samples a \textit{seed core node} $v_s\in \mathcal{C}$ according to a probability distribution $\vect{p}_\text{seed}\in [0,1]^{|\mathcal{C}|}$, where $\|\vect{p}_\text{seed}\|_1=1$,
    and $\vect{p}_\text{seed}(v)$ is the probability of selecting node $v\in \mathcal{C}$ 
    \blue{as a seed core}.
    Then, \method calculates the inclusion probability of remaining core node $v_c \in \mathcal{C} \setminus \{v_s\}$ to the core group based on its attribute affinity with the seed node $v_s$, computed as:
    \begin{equation}\label{eq:p_g}
        P_{\mathcal{C}}(v_c | v_s, \Theta_{\mathcal{C}}) = 
        \prod\nolimits_{l=1}^k \mat{\theta}_{\mathcal{C}}^{(l)}[\mathbf{x}_s^{(l)}, \mathbf{x}_c^{(l)}],\vspace{-1mm}
    \end{equation}
    where $\Theta_{\mathcal{C}} = \{\mat{\theta}_{\mathcal{C}}^{(1)}, \dots, \mat{\theta}_{\mathcal{C}}^{(k)}\}$ is the set of attribute affinity matrices modeling interactions \textit{among core nodes}, and each $\mat{\theta}_{\mathcal{C}}^{(l)}\in \mathbb{R}^{2\times 2}$ captures the affinity between binary values of the $l$-th attribute. 
    \item \textbf{Step 2. Fringe Attachment (lines \ref{alg:method:fringestart} -- \ref{alg:method:fringeend}):}
    Subsequently, \method samples fringe nodes from $\mathcal{F}$ to attach to the core group $\mathcal{C}_{\tilde{e}}$ and complete the hyperedge construction.
    For each fringe node $v_f\in \mathcal{F}$, \method first constructs a binary attribute vector $\mathbf{x}_{\mathcal{C}_{\tilde{e}}} \in \{0,1\}^k$ that summarizes the attributes of the nodes in the core group $\mathbf{x}_{\mathcal{C}_{\tilde{e}}}$.
    Each $l$-th attribute $\mathbf{x}_{\mathcal{C}_{\tilde{e}}}^{(l)}$ is independently sampled as:
    \begin{equation} \label{eq:attribute_sampling}
    \mathbf{x}_{{\mathcal{C}_{\tilde{e}}}}^{(l)} \sim \text{Bernoulli}\left( \frac{1}{\vert {\mathcal{C}_{\tilde{e}}} \vert} \sum\nolimits_{v_i \in {\mathcal{C}_{\tilde{e}}}} \mathbf{x}_i^{(l)} \right), \vspace{-1mm}
    \end{equation}
    which intuitively, samples the attribute according to its average presence among the group members.
    Then, $v_f$ is considered for attachment with the following probability:
    \begin{equation}\label{eq:p_f}
    P_{\mathcal{F}}(v_f | \mathcal{C}_{\tilde{e}}, \Theta_{\mathcal{F}}) = \prod\nolimits_{l=1}^k \mat{\theta}_{\mathcal{F}}^{(l)}[\mathbf{x}_{\mathcal{C}_{\tilde{e}}}^{(l)}, \mathbf{x}_f^{(l)}], \vspace{-1mm}
    \end{equation}
    where $\Theta_{\mathcal{F}} = \{\mat{\theta}_{\mathcal{F}}^{(1)}, \dots, \mat{\theta}_{\mathcal{F}}^{(k)}\}$ is the set of attribute affinity matrices that model interactions \textit{between the core group and fringe nodes}, and each $\mat{\theta}_{\mathcal{F}}^{(l)}\in \mathbb{R}^{2\times 2}$ captures the affinity between binary values of the $l$-th attribute. 
\end{itemize}

\subsection{Theoretical Analysis of \method.}
We analyze complexity and structural properties of \method.

\smallsection{Complexity.} 
For core group construction, sampling the seed core node $v_s$ takes $O(\vert \mathcal{C} \vert)$ time, while computing attachment probabilities $P_\mathcal{C}$ for the remaining core nodes $\mathcal{C}\setminus \{v_s\}$ requires $O(k \vert \mathcal{C} \vert)$ time. 
For fringe attachment, computing the attachment probabilities $P_\mathcal{F}$ for fringe nodes $\mathcal{F}$ takes $O(k \vert \mathcal{F} \vert)$ time.
\method generates a hypergraph $\Tilde{\mathcal{H}}$ consisting of $m$ hyperedges by repeating these two steps $m$ times.
Thus, the overall time complexity of the hypergraph generation process of \method is $O(mk |\mathcal{V}|)$ where $|\mathcal{V}| = |\mathcal{C}| + |\mathcal{F}|$.
\red{Regarding space complexity, \method requires $O(1)$ space per hyperedge to store the seed core.}
The sampled attribute of a core group requires $O(k)$ space, which can be discarded after computing the attachment probabilities for the fringe nodes.
Thus, the overall space complexity of the hypergraph generation by \method is $O(k)$.

\smallsection{Structural Properties.} \method is capable of generating hypergraphs with heavy-tailed node degree distributions, which is common in real-world hypergraphs~\cite{ko2022growth,lee2024survey}.
\begin{theorem} \label{thm:power-law-degree}
    There exist configurations on node attributes and parameters on \method such that the generated hypergraph follows a power-law degree distribution.
\end{theorem}
\noindent
Proof of Theorem \ref{thm:power-law-degree} is in Appendix VIII~\cite{online}.

\begin{algorithm}[!t]
\small
\caption{\method}\label{alg:method}

\nonl \textbf{Input:} (1) number of hyperedges $m$\\
\nonl \hspace{0.86cm} (2) node attribute matrix $\mat{X} \in {\{0,1\}}^{\vert \mathcal{V} \vert \times k}$ \\
\nonl \hspace{0.86cm} (3) set of core nodes $\mathcal{C}$ \\
\nonl \hspace{0.86cm} (4) set of fringe nodes $\mathcal{F}$\\
\nonl \hspace{0.86cm} (5) seed core probabilities $\vect{p_\text{seed}}$ \\
\nonl \hspace{0.86cm} (6) set of core group affinity matrices $\Theta_{\mathcal{C}}$\\
\nonl \hspace{0.86cm} (7) set of fringe attachment affinity matrices $\Theta_{\mathcal{F}}$\\

\nonl \textbf{Output:} generated hypergraph $\Tilde{\mathcal{H}} = (\mathcal{V}, \Tilde{\mathcal{E}})$

$\Tilde{\mathcal{H}} = (\mathcal{V}, \Tilde{\mathcal{E}}=\varnothing)$


\For{\textbf{each} $i=1,\cdots,m$}{
    \tcp{\algblue{1. Core Group Construction}}

    Sample $v_s \sim \vect{p}_{\text{seed}}$, $v_s\in \mathcal{C}$ \label{alg:method:cgstart}
    
    $\mathcal{C}_{\tilde{e}} \leftarrow \{v_s\}$   
    
    \For{\textbf{each} $v_c \in \mathcal{C} \setminus \{v_s\}$ }{
        $p \leftarrow P_{\mathcal{C}}(v_c | v_s, \Theta_{\mathcal{C}})$ \algblueline{\texttt{Eq.~\eqref{eq:p_g}}}
        
        \textbf{with} \textbf{\textit{probability}} $p$ \textbf{do} 
        
        $\;\;\;\;\;\;\mathcal{C}_{\tilde{e}} \leftarrow \mathcal{C}_{\tilde{e}} \cup \{v_c\}$ \label{alg:method:cgend}\label{alg:method:coreadd2}
    }
    

    \tcp{\algblue{2. Fringe Node Attachment}}
    $\mathcal{F}_{\tilde{e}} \leftarrow \varnothing$\label{alg:method:fringestart}
    
    \For{\textbf{each} $v_f \in \mathcal{F}$}{ 
        
        

        Construct $\mathbf{x}_{\mathcal{C}_{\tilde{e}}}$ of $\mathcal{C}_{\tilde{e}}$ \algblueline{\texttt{Eq.~\eqref{eq:attribute_sampling}}}

        $q \leftarrow P_{\mathcal{F}}(v_f | \mathcal{C}_{\tilde{e}}, \Theta_{\mathcal{F}})$ \algblueline{\texttt{Eq.~\eqref{eq:p_f}}}
        
        \textbf{with} \textbf{\textit{probability}} $q$ \textbf{do}
        
        $\;\;\;\;\;\;\mathcal{F}_{\tilde{e}} \leftarrow \mathcal{F}_{\tilde{e}} \cup \{v_f\}$ \label{alg:method:fringeend}
    }

    $\tilde{e}\leftarrow \mathcal{C}_{\tilde{e}} \cup \mathcal{F}_{\tilde{e}}$
    
    $\Tilde{\mathcal{E}} \leftarrow \Tilde{\mathcal{E}} \cup \{\Tilde{e}\}$
}
\Return $\Tilde{\mathcal{H}} = (\mathcal{V}, \Tilde{\mathcal{E}})$
\end{algorithm}

%% file: 050fitting.tex
In this section, we propose \fitwattr, a method for tuning the parameters of \method to fit a given hypergraph, outlined in Algorithm~\ref{alg:fit}.
The goal of fitting is to enable \method to generate hypergraphs that closely resemble the input in both structure and attribute patterns. \red{This is useful for} various downstream applications, such as data anonymization and simulation, where generating realistic yet controllable hypergraphs is critical. 
We begin by describing 
how the nodes $\mathcal{V}$ is partitioned into core nodes $\mathcal{C}$ and fringe nodes $\mathcal{F}$.
\red{Next, we derive hyperedge likelihoods, which are a key component for parameter optimization.
Then, we discuss parameter optimization for maximizing the likelihood of the given hypergraph.
Finally, we analyze the complexity of \fitwattr.} 

\subsection{Core and Fringe Partition} \label{sec:fitting:node_dividing}
Given a hypergraph $\mathcal{H}$, \fitwattr partitions the node set $\mathcal{V}$ into a core node set $\mathcal{C}$ and a fringe node set $\mathcal{F}$.
We utilize UMHS (see Section~\ref{sec:prelim:umhs}) to identify the core node set $\mathcal{C}$.
Then, the fringe set is defined as its complement, i.e., $\mathcal{F} = \mathcal{V} \setminus \mathcal{C}$.
Each hyperedge $e\in \mathcal{E}$ consists of a subset of core nodes $\mathcal{C}_e = e \cap \mathcal{C}$ and a subset of fringe nodes $\mathcal{F}_e = e \cap \mathcal{F}$.

\subsection{Derivation of the Likelihood of Each Hyperedge} \label{sec:fitting:likelihood}
Given the structural and attribute information of the hypergraph (i.e., $\mathcal{C}$, $\mathcal{F}$, and $\mathbf{X}$) and the parameters of \method (i.e., $\vect{p}_\text{seed}$, $\Theta_\mathcal{C}$, and $\Theta_\mathcal{F}$), we now derive the likelihood of each hyperedge $e\in \mathcal{E}$, denoted as $P(e | \mathcal{C}, \mathcal{F}, \mathbf{X}, \vect{p}_\text{seed}, \Theta_\mathcal{C}, \Theta_\mathcal{F})$, which we refer to as $P(e)$ for brevity.
The likelihood is decomposed into two components: 
(1) $P_\text{core}(e)$, the likelihood of core group construction, and 
(2) $P_\text{fringe}(e)$, the likelihood of fringe node attachment.
The total likelihood is then given by $P(e)=P_\text{core}(e)\cdot P_\text{fringe}(e)$.

\smallsection{Likelihood of Sampling Core Nodes.} 
\red{Given a hyperedge $e$, the likelihood $P_\text{core}(e)$ of its core group $\mathcal{C}_e$ is:} 
\begin{equation}\label{eq:p_core}
    P_\text{core}(e) = \sum\nolimits_{v_s \in \mathcal{C}_e} \vect{p}_\text{seed}(v_s) \cdot P_\mathcal{C}(\mathcal{C}_e \setminus \{v_s\} | v_s),
    ~\footnote{For training stability, in practice, we implement \fitwattr by using a normalized core group likelihood, where $P_\text{core}(e)$ is divided by the total seed probability over the core group, i.e., $P_\text{core}(e) / \sum_{v_c \in \mathcal{C}_e}\vect{p}_\text{seed}(v_c)$.}
\end{equation}
where $P_\mathcal{C}(\mathcal{C}_e \setminus \{v_s\} | v_s)$ is the likelihood of sampling the remaining core nodes $\mathcal{C}_e \setminus \{v_s\}\subset \mathcal{C}$ given the seed node $v_s$, which can be written as:
\begin{equation*}
    P_\mathcal{C}(\mathcal{C}_e \setminus \{v_s\} | v_s) = \prod_{v_c\in \mathcal{C}_e \setminus \{v_s\}} P_\mathcal{C}(v_c | v_s)\cdot \prod_{v_c \in \mathcal{C} \setminus \mathcal{C}_e}(1 - P_\mathcal{C}(v_c | v_s)).
\end{equation*}
For brevity, we omit $\Theta_\mathcal{C}$ in the above equations. 


    
\smallsection{Likelihood of Sampling Fringe Nodes.}
Given the core group $\mathcal{C}_e$ of the hyperedge $e$, the likelihood of its fringe subset $\mathcal{F}_e$ is:
\begin{equation}\label{eq:p_fringe}
    P_\text{fringe}(e) = \prod_{v_f\in \mathcal{F}_e} P_\mathcal{F}(v_f | \mathcal{C}_e) \cdot \prod_{v_f \in \mathcal{F} \setminus \mathcal{F}_e} (1 - P_\mathcal{F}(v_f | \mathcal{C}_e)),
\end{equation}
where $P_\mathcal{F}(v_f | \mathcal{C}_e)$ is the probability of attaching fringe node $v_f$ to the core group $\mathcal{C}_e$.
This probability is obtained by marginalizing over the stochastic binary attribute vector $\mathbf{x}_{\mathcal{C}_e}$ of $\mathcal{C}_e$ as follows:
\begin{align*}
    P_\mathcal{F}(v_f | \mathcal{C}_e) &= \mathbb{E}_{\mathbf{x}_{\mathcal{C}_e}}\left[ \prod\nolimits_{l=1}^k \mat{\theta}_\mathcal{F}^{(l)}[\mathbf{x}_{\mathcal{C}_e}^{(l)}, \mathbf{x}_f^{(l)}] \right]\\
    &= \prod_{l=1}^k \left[ (1 - p_e^{(l)})\cdot \mat{\theta}_\mathcal{F}^{(l)}[0,\mathbf{x}_f^{(l)}] + p_e^{(l)} \cdot \mat{\theta}_\mathcal{F}^{(l)}[1,\mathbf{x}_f^{(l)}] \right],
\end{align*}
where each $l$-th component of $\mathbf{x}_{\mathcal{C}_e}$ is sampled as a Bernoulli random variable with mean equal to $p_e^{(l)}=\frac{1}{|\mathcal{C}_e|}\sum_{v_i\in \mathcal{C}_e}\mathbf{x}_i^{(l)}$.



\subsection{Update of the Parameters of \method} \label{sec:fitting:loss}
We present the loss functions used to update the parameters of \method (spec., $\vect{p}_\text{seed}$, $\Theta_\mathcal{C}$, and $\Theta_\mathcal{F}$) as follows:
\begin{itemize}[leftmargin=*]
    \item \textbf{Negative Log-likelihood Loss ($\mathcal{L}_{edge}$):} For each hyperedge $e$, we calculate the negative log-likelihood, i.e., $-\log P(e)$.  
These values are then summed over all hyperedges to compute the negative log-likelihood loss:
\begin{align*} \label{eq:likelihood_fringe}
\\[-6.5mm]
\mathcal{L}_{\text{edge}} = \sum\nolimits_{e \in \mathcal{E}} -\log P(e). \\[-6.5mm]
\end{align*}
    \item \textbf{Degree and Cardinality Losses ($\mathcal{L}_{deg}$  and $\mathcal{L}_{card}$):}
    Unlike many existing hypergraph models that rely on degree or cardinality distributions as explicit model inputs, \method  learns to reproduce \red{realistic degree and cardinality patterns, without using them as model inputs (only using them for fitting), 
    demonstrating its expressive modeling capability.}
    To encourage this behavior, we use a mean squared error (MSE) loss that aligns the expected degrees and hyperedge sizes with the distributions in the original hypergraph:
    \begin{align*}
    \\[-6.5mm]
    \mathcal{L}_{deg} = \textsc{MSE}(\vect{d}_c, \tilde{\vect{d}_c})+\textsc{MSE}(\vect{d}_f, \tilde{\vect{d}_f})\\
    \mathcal{L}_{card} = \textsc{MSE}(\vect{c}_c, \tilde{\vect{c}_c})+\textsc{MSE}(\vect{c}_f, \tilde{\vect{c}_f}), \\[-6.5mm]
    \end{align*}
    where $\vect{d}_c, \vect{d}_f$ denote the degrees of core and fringe nodes; and $\vect{c}_c ,\vect{c}_f$ denote the \red{cardinalities of core and fringe subsets within hyperedges.} 
    \red{The corresponding quantities with tildes indicate their expected values under \method.
    We compare distributions by sorting them and computing the MSE between corresponding values, focusing on overall distributional similarity rather than individual node identities.}
\end{itemize}
\red{The final loss is a weighted sum of the above losses (see Line~\ref{alg:fit:lossend} of Algorithm~\ref{alg:fit}), with the weights as hyperparameters.}

\subsection{Complexity of \fitwattr.}
We provide the time and space complexity analysis of \fitwattr.
Regarding the time complexity, the core and fringe partitioning \cite{amburg2021planted} requires $O(m\vert \mathcal{V} \vert)$ time.
For each hyperedge $e\in \mathcal{E}$, computing its likelihood $P(e)$ takes $O(k  |\mathcal{V}| |\mathcal{C}_e|)$ time.
With attachment probabilities calculated during the computation of $P(e)$, calculation of expected degree and cardinality from \method takes $O(\vert \mathcal{V} \vert)$ time.
Repeating this for $T$ training epochs over all hyperedges results in a total computation cost of $O(Tk |\mathcal{V}| \sum_{e\in\mathcal{E}}|\mathcal{C}_e|)$.
Regarding the space complexity, 
\blue{
core and fringe partition takes $O(\vert \mathcal{V} \vert)$ space.
For each hyperedge $e \in \mathcal{E}$, computing its likelihood $P(e)$ takes $O(k + \vert \mathcal{V} \vert)$ space.
Expected degree and cardinality require $O(m + \vert \mathcal{V} \vert)$ space.
Thus, the total space complexity is $O(k + m + \vert \mathcal{V} \vert)$.
}

\begin{algorithm}[!t]
\small
\caption{\fitwattr}\label{alg:fit}
\nonl \textbf{Input:} (1) target hypergraph $\mathcal{H}=(\mathcal{V},\mathcal{E},\mat{X})$ \\
\nonl \hspace{0.86cm} (2) number of epochs $T$, learning rate $\eta$ \\
\nonl \hspace{0.86cm} (3) loss weights $w_{deg}, w_{card}$ \\
\nonl \textbf{Output:} (1) set of core fringe nodes $\mathcal{C}$ \\
\nonl \hspace{1.09cm} (2) set of core fringe nodes $\mathcal{F}$ \\
\nonl \hspace{1.09cm} (3) seed core probabilities $\vect{p_{\text{seed}}}$ \\
\nonl \hspace{1.09cm} (4) set of core group affinity matrices $\Theta_{\mathcal{C}}$ \\
\nonl \hspace{1.09cm} (5) set of fringe attachment affinity matrices $\Theta_{\mathcal{F}}$ \\

\tcp{\algblue{1. Core-fringe Split}} 

Split $\mathcal{V}$ into $\mathcal{C}, \mathcal{F}$

Initialize $\vect{p_{\text{seed}}},\Theta_{\mathcal{C}}$ and $\Theta_{\mathcal{F}}$~\footnotemark

\For{\textbf{each} $t=1,\cdots,T$}{
    \tcp{\algblue{2. Log-likelihood Loss}}
    
    $\mathcal{L}_{edge} \leftarrow 0$ 
    
    \For{\textbf{each} $e \in \mathcal{E}$}{
        $\mathcal{C}_e\leftarrow e\cap\mathcal{C},\;\mathcal{F}_e\leftarrow e\cap\mathcal{F}$

        $P(e) \leftarrow P_\text{core}(e) \cdot P_\text{fringe}(e)$ 
        \algblueline{\texttt{Eq.~\eqref{eq:p_core} \& Eq.~\eqref{eq:p_fringe}}}
        
        $\mathcal{L}_{edge}\leftarrow \mathcal{L}_{edge} - \log P(e)$   
        
    }

    \tcp{\algblue{3. Degree, Cardinality Losses}}
    
    
    Compute $\tilde{\vect{d}_c}, \tilde{\vect{d}_f}$ and $\tilde{\vect{c}_c}, \tilde{\vect{c}_f}$ using $\vect{p}_{\text{seed}},\Theta_{\mathcal{C}}$ and $\Theta_{\mathcal{F}}$

    $\mathcal{L}_{deg} \leftarrow \textsc{MSE}(\vect{d}_c, \tilde{\vect{d}_c})+\textsc{MSE}(\vect{d}_f, \tilde{\vect{d}_f})$

    $\mathcal{L}_{card} \leftarrow \textsc{MSE}(\vect{c}_c, \tilde{\vect{c}_c})+\textsc{MSE}(\vect{c}_f, \tilde{\vect{c}_f})$

    $\mathcal{L} \leftarrow \mathcal{L}_{edge} + w_{deg} \cdot \mathcal{L}_{deg} + w_{card} \cdot \mathcal{L}_{card}$ \label{alg:fit:lossend}

    \tcp{\algblue{4. Parameter Update}}

    $\vect{p_\text{seed}}\leftarrow\vect{p_\text{seed}}+ \eta \nabla_{\vect{p_\text{seed}}} \mathcal{L}$
    
    $\Theta_\mathcal{C} \leftarrow \Theta_\mathcal{C} + \eta \nabla_{\Theta_\mathcal{C}} \mathcal{L}$
    
    $\Theta_\mathcal{F} \leftarrow \Theta_\mathcal{F} + \eta \nabla_{\Theta_\mathcal{F}} \mathcal{L}$
}
\KwRet $\mathcal{C}$, $\mathcal{F}$, $\vect{p_{\text{seed}}}$, $\Theta_{\mathcal{C}}$, $\Theta_{\mathcal{F}}$
\end{algorithm}
\footnotetext{Initialization method of parameters is explained in Appendix IX~\cite{online}.}

%% file: 060exp.tex
In this section, we present experimental results demonstrating the effectiveness of \method and \fitwattr.

\subsection{Experimental Settings}

\smallsection{Datasets.}
We use nine real-world hypergraphs from four distinct domains (see Table \ref{tab:datastat} for some statistics):
\begin{itemize}[leftmargin=*]
\item \textbf{Academic Paper Domain (Citeseer, Cora~\cite{yadati2019hypergcn}):} Each node is an academic paper. For the Citeseer dataset, each hyperedge is a set of papers co-cited by a paper, and for the Cora dataset, each hyperedge is a set of papers (co-)authored by the same author. \red{Node attributes are binary bag-of-words, indicating whether a paper includes each keyword or not.}
\item \textbf{Contact domain (High School~\cite{chodrow2021hypergraph}, Workspace~\cite{genois2018can}):} Each node is an individual, and each hyperedge is a group of individuals in contact during a time interval. For the High School dataset, \red{node attributes include} gender, class affiliation, and Facebook account ownership. For the Workspace dataset, node attributes represent the department of each worker. Since all attributes are categorical, we apply one-hot encoding to convert them into binary attributes.
\item \textbf{Review Domain (Amazon Music~\cite{ni2019justifying}, Yelp Restaurant, Yelp Bar~\cite{amburg2020fair}):} Each node is a reviewer, and each hyperedge is a group of reviewers who reviewed a certain product or \red{business}. Node attributes indicate the types of products or businesses that each reviewer has reviewed at least once.
\item \textbf{Online Q\&A Domain\footnote {https://archive.org/download/stackexchange}(Devops, Patents):} Each node represents a user on Stack Exchange, and each hyperedge corresponds to a post involving a set of users. \red{Node attributes indicate the set of tags associated with the posts each user has participated in.}
\end{itemize}
Note that our experiments are done with \red{varying numbers of attributes,} scaling from 5 (Workspace) to 3,703 (Citeseer).

\setlength{\tabcolsep}{4pt} 
\begin{table}[t]
    \vspace{-2mm}
    \centering
    \caption{\centering Summary statistics of 9 real-world hypergraphs from 4 domains. $\vert \mathcal{V} \vert$: the number of nodes. $\vert \mathcal{E} \vert$: the number of hyperedges. $k$: the number of attributes.
    The core-set size $\vert \mathcal{C} \vert$ and the fringe-set size $\vert \mathcal{F} \vert$ are obtained by \fitwattr.
    } 
    \label{tab:datastat}
    \begin{tabular}{p{2.7cm}||Y{0.7cm}|Y{0.7cm}|Y{0.7cm}||Y{0.7cm}|Y{0.7cm}}
        \toprule
        \textbf{Dataset} & $\vert \mathcal{V} \vert$ & $\vert \mathcal{E} \vert$ & $k$ & $\vert \mathcal{C} \vert$ & $\vert \mathcal{F} \vert$ \\
        \midrule
        Citeseer & 1,458 & 1,079 & 3,703 & 597 & 861 \\
        Cora & 2,388 & 1,072 & 1,433 & 841 & 1,547 \\ 
        \midrule
        High School & 327 & 7,818 & 12 & 288 & 39 \\
        Workspace & 92 & 788 & 5 & 71 & 21 \\
        \midrule
        Amazon Music & 1,106 & 686 & 7 & 379 & 727 \\
        Yelp Restaurant & 565 & 594 & 9 & 273 & 292 \\
        Yelp Bar & 1,234 & 1,188 & 15 & 625 & 609 \\
        \midrule
        Devops & 5,010 & 5,684 & 429 & 2,003 & 3,007 \\
        Patents & 4,458 & 4,669 & 2,170 & 894 & 3,564 \\
        \bottomrule
    \end{tabular}
\end{table}

\begin{table*}[!htbp]
\vspace{-5mm}
\centering
\caption{\method reproduces structure-attribute interplays overall best across $9$ datasets. Top three results are highlighted in \textcolor{rankone}{blue} (first), \textcolor{ranktwo}{green} (second), and \textcolor{textyellow}{yellow} (third). 
Refer to Section~\ref{sec:exp:ablation} for 
\method-CF, a variant of \method.
A.R. denotes average rank.
}
\label{tab:interplayeval}
\vspace{-2mm}
\subfloat[Citeseer (\method ranks \textbf{first} overall)]{
\begin{tabularx}{\columnwidth}{p{1.9cm}|Y{0.7cm}Y{0.7cm}Y{0.7cm}Y{0.7cm}Y{0.7cm}Y{0.7cm}|Y{0.5cm}}
    \toprule
    & \textbf{T2} & \textbf{T3} & \textbf{T4} & \textbf{HE} & \textbf{HOHE} & \textbf{NHS} & \textbf{A.R.} \\
    \midrule
    \hypercl & 6,816 & 10,702 & 10,672 & 19.81 & 122.02 & 19.94 & 6.7 \\
    \hyperpa & 6,871 & 10,739 & 10,559 & 25.90 & 103.76 & 22.46 & 7.7 \\
    \hyperff & 6,472 &\cellcolor{rankthree}10,483 & 10,677 &\cellcolor{ranktwo}17.60 & 66.97 &\cellcolor{rankone}15.23 & 4.0 \\
    \hyperlap & 6,757 & 10,737 & 10,311 & 18.33 & 55.40 & 19.43 & 5.0 \\
    \hyperdk & 6,968 &\cellcolor{ranktwo}10,234 &\cellcolor{ranktwo}10,154 &\cellcolor{rankone}13.38 & 102.83 &\cellcolor{rankthree}16.09 &\cellcolor{ranktwo}3.7 \\
    \thera &\cellcolor{rankthree}6,450 & 10,498 &10,476 &\cellcolor{rankthree}18.04 & \cellcolor{rankthree}53.58 & 18.59 &\cellcolor{rankthree}3.8\\
    \hycosbm &\cellcolor{ranktwo}6,285 &10,613 & 10,278 & 19.89 & 135.30 & 38.60 & 6.0 \\
    \hyrec & 6,996 & 10,840 &\cellcolor{rankthree}10,271 & 20.01 &\cellcolor{ranktwo}51.37 & 20.04 & 6.2 \\
    \midrule
    \textbf{\method} &\cellcolor{rankone}5,734 &\cellcolor{rankone}10,157 &\cellcolor{rankone} 10,151 & 24.31 &\cellcolor{rankone}44.26 &\cellcolor{ranktwo}15.39 &\cellcolor{rankone}2.3 \\
    \textbf{\method-CF} & 9,036 & 14,550 & 13,515 & 48.26 & 106.32 & 121.04 & 9.7 \\
    \bottomrule
\end{tabularx}
}
\hfill
\subfloat[Cora (\method ranks \textbf{second} overall)]{
\begin{tabularx}{\columnwidth}{p{1.9cm}|Y{0.7cm}Y{0.7cm}Y{0.7cm}Y{0.7cm}Y{0.7cm}Y{0.7cm}|Y{0.5cm}}
    \toprule
    & \textbf{T2} & \textbf{T3} & \textbf{T4} & \textbf{HE} & \textbf{HOHE} & \textbf{NHS} & \textbf{A.R.}\\
    \midrule
    \hypercl & 2,099 & 4,206 &\cellcolor{ranktwo}4,481 & 8.02 & 63.64 & 7.99 & 7.0 \\
    \hyperpa & 2,081 & 4,315 & 4,615 & 14.47 & 62.62 & 12.78 & 8.3 \\
    \hyperff & 2,040 & 4,094 & 4,518 & 8.06 & \cellcolor{rankthree}39.58 & \cellcolor{rankthree}6.76 & 5.0 \\
    \hyperlap &\cellcolor{rankthree}2,031 & 4,083 & 4,523 & 7.65 & 53.40 & 7.13 & 5.2 \\
    \hyperdk & 2,094 &\cellcolor{rankthree}4,048 & 4,502 & 7.24 & 59.79 & 7.49 & 5.5 \\
    \thera &\cellcolor{ranktwo}2,010 &\cellcolor{rankone}4,003 &\cellcolor{rankthree}4,492 &\cellcolor{rankthree}6.94 & \cellcolor{ranktwo}40.35 & 7.20 &\cellcolor{rankone}2.8 \\
    \hycosbm &\cellcolor{rankone}1,947 & 4,049 & 4,502 & 7.24 & 69.77 & 15.02 & 5.5 \\
    \hyrec & 2,075 & 4,163 &\cellcolor{rankone}4,465 &\cellcolor{ranktwo}6.51 & \cellcolor{rankone}19.56 & \cellcolor{ranktwo}6.54 & \cellcolor{ranktwo}3.2 \\
    \midrule
    \textbf{\method} & 2,058 &\cellcolor{ranktwo}4,011 & 4,517 & \cellcolor{rankone}6.41 & 41.64 &\cellcolor{rankone}5.79 & \cellcolor{ranktwo}3.2 \\
    \textbf{\method-CF} & 3,322 & 5,951 & 5,995 & 29.01 & 57.35 & 70.15 & 9.3 \\
    \bottomrule
\end{tabularx}
}
\vspace{0.1mm}
\subfloat[High School (\method ranks \textbf{first} overall)]{
\begin{tabularx}{\columnwidth}{p{1.9cm}|Y{0.7cm}Y{0.7cm}Y{0.7cm}Y{0.7cm}Y{0.7cm}Y{0.7cm}|Y{0.5cm}}
    \toprule
    & \textbf{T2} & \textbf{T3} & \textbf{T4} & \textbf{HE} & \textbf{HOHE} & \textbf{NHS} & \textbf{A.R.} \\
    \midrule
    \hypercl   & 20.4 & 51.4 & 106.4 & 1.180 & 1.369 & 1.730 & 7.2 \\
    \hyperpa   & 20.6 & 52.6 & 107.8 & 1.175 & 1.421 & 1.654 & 8.0 \\
    \hyperff   & 20.2 & 61.6 & 106.5 & 1.017 & 1.290 & 1.785 & 7.8 \\
    \hyperlap   & 20.3 & 51.6 & 102.0 & 1.187 & \cellcolor{rankthree}0.977 & 1.714 & 6.0 \\
    \hyperdk   & 20.0 & 51.4 & \cellcolor{rankthree}97.8 & 0.931 & 1.429 & 1.716 & 5.8 \\
    \thera   & 19.9 & \cellcolor{rankthree}51.2 & 99.8 & 1.166 & \cellcolor{rankone}0.781 & \cellcolor{ranktwo}1.611 & \cellcolor{ranktwo}3.5 \\
    \hycosbm   & \cellcolor{rankone}3.9 & 52.1 & 97.9 & 1.819 & 1.414 & 1.797 & 6.7 \\
    \hyrec   & 19.8 & 54.0 & \cellcolor{ranktwo}92.8 & \cellcolor{rankthree}0.740 & \cellcolor{ranktwo}0.960 & 1.637 & 4.3 \\
    \midrule
    \textbf{\method} & \cellcolor{ranktwo}12.2 & \cellcolor{rankone}37.8 & \cellcolor{rankone}89.6 & \cellcolor{rankone}0.628 & 1.273 & \cellcolor{rankone}1.374 & \cellcolor{rankone}2.0 \\
    \textbf{\method-CF} & \cellcolor{rankthree} 19.2 &  \cellcolor{ranktwo}50.3 & 103.9 & \cellcolor{ranktwo} 0.682 & 1.176 &\cellcolor{rankthree}1.635 & \cellcolor{rankthree}3.7 \\
    \bottomrule
\end{tabularx}
}
\hfill
\subfloat[Workspace (\method ranks \textbf{first} overall)]{
\begin{tabularx}{\columnwidth}{p{1.9cm}|Y{0.7cm}Y{0.7cm}Y{0.7cm}Y{0.7cm}Y{0.7cm}Y{0.7cm}|Y{0.5cm}}
    \toprule
    & \textbf{T2} & \textbf{T3} & \textbf{T4} & \textbf{HE} & \textbf{HOHE} & \textbf{NHS} & \textbf{A.R.} \\
    \midrule
    \hypercl   & 7.5 & 17.3 & 13.0 & 0.599 & 0.181 & 0.890 & 7.0 \\
    \hyperpa   & 7.1 & 19.5 & \cellcolor{rankone}12.5 & 0.562 & 0.217 & 0.835 & 6.0 \\
    \hyperff   & 6.3 & 14.9 & 13.9 & 0.445 & 0.399 & \cellcolor{rankthree}0.801 & \cellcolor{rankthree}5.2 \\
    \hyperlap  & 7.3 & 19.9 & \cellcolor{rankthree}12.6 & 0.564 & 0.184 & 0.878 & 7.0 \\
    \hyperdk   & 6.5 & 15.6 & 25.1 & 0.408 & \cellcolor{rankthree}0.178 & 0.814 & \cellcolor{rankthree}5.2 \\
    \thera     & \cellcolor{rankthree}5.7 & \cellcolor{rankthree}13.7 & 16.7 & 0.542 & \cellcolor{ranktwo}0.144 & 0.820 & \cellcolor{ranktwo}4.2 \\
    \hycosbm   & \cellcolor{rankone}2.2 & \cellcolor{ranktwo}13.0 & 24.0 & 0.705 & 0.233 & 0.885 & 6.2 \\
    \hyrec     & 7.3 & 18.0 & 20.3 & \cellcolor{rankthree}0.371 & 0.301 & \cellcolor{ranktwo}0.753 & 6.2 \\
    \midrule
    \textbf{\method}  & \cellcolor{ranktwo}4.2 & \cellcolor{rankone}9.7 & 20.0 & \cellcolor{rankone}0.062 & \cellcolor{rankone}0.133 & \cellcolor{rankone}0.619 & \cellcolor{rankone}2.2 \\
    \textbf{\method-CF} & 7.3 & 18.4 & \cellcolor{rankone}12.5 & \cellcolor{ranktwo}0.186 & 0.283 & 0.905 & 6.0 \\
    \bottomrule
\end{tabularx}
}
\vspace{0.1mm}
\subfloat[Amazon Music (\method ranks \textbf{first} overall)]{
\begin{tabularx}{\columnwidth}{p{1.9cm}|Y{0.7cm}Y{0.7cm}Y{0.7cm}Y{0.7cm}Y{0.7cm}Y{0.7cm}|Y{0.5cm}}
    \toprule
    & \textbf{T2} & \textbf{T3} & \textbf{T4} & \textbf{HE} & \textbf{HOHE} & \textbf{NHS} & \textbf{A.R.} \\
    \midrule
    \hypercl   & 27.3 & 53.0 & 63.6 & 1.016 & 0.382 & 1.053 & 6.8 \\
    \hyperpa   & 27.3 & 55.4 & 71.2 & 1.154 & 0.527 & 1.096 & 8.8 \\
    \hyperff   & 24.1 & 54.3 & \cellcolor{rankthree}60.5 & 0.449 & \cellcolor{rankone}0.299 & 1.055 & 5.2 \\
    \hyperlap  & 27.3 & 52.4 & 68.3 & 1.026 & \cellcolor{ranktwo}0.361 & 1.042 & 6.3 \\
    \hyperdk   & 31.4 & 52.4 & 61.6 & 1.249 & 0.450 & 1.026 & 7.3 \\
    \thera     & 26.0 & \cellcolor{rankthree}50.6 & 67.0 & 0.976 & 0.394 & 1.003 & 5.0 \\
    \hycosbm   & \cellcolor{rankone}11.8 & 57.9 & 72.1 & \cellcolor{ranktwo}0.306 & \cellcolor{rankthree}0.371 & \cellcolor{rankthree}0.900 & \cellcolor{rankthree}4.7 \\
    \hyrec     & 25.3 & \cellcolor{rankthree}50.6 & 61.8 & 1.138 & 0.402 & 0.982 & 5.3 \\
    \midrule
    \textbf{\method}    & \cellcolor{ranktwo}21.0 & \cellcolor{rankone}47.8 & \cellcolor{rankone}55.1 & \cellcolor{rankone}0.275 & 0.394 & \cellcolor{rankone}0.229 & \cellcolor{rankone}1.8 \\
    \textbf{\method-CF} & \cellcolor{rankthree}21.8 & \cellcolor{ranktwo}49.7 & \cellcolor{ranktwo}58.0 & \cellcolor{rankthree}0.363 & 1.188 & \cellcolor{ranktwo}0.402 & \cellcolor{ranktwo}3.7 \\
    \bottomrule
\end{tabularx}
}
\hfill
\subfloat[Yelp Restaurant (\method ranks \textbf{second} overall)]{
\begin{tabularx}{\columnwidth}{p{1.9cm}|Y{0.7cm}Y{0.7cm}Y{0.7cm}Y{0.7cm}Y{0.7cm}Y{0.7cm}|Y{0.5cm}}
    \toprule
    & \textbf{T2} & \textbf{T3} & \textbf{T4} & \textbf{HE} & \textbf{HOHE} & \textbf{NHS} & \textbf{A.R.} \\
    \midrule
    \hypercl   & \cellcolor{rankthree}17.8 & 58.9 & 90.4 & 0.979 & \cellcolor{ranktwo}0.249 & 1.148 & 6.5 \\
    \hyperpa   & 24.8 & 55.9 & \cellcolor{ranktwo}76.0 & 0.786 & 0.591 & 1.068 & 6.3 \\
    \hyperff   & 22.4 & \cellcolor{rankthree}54.1 & 79.2 & \cellcolor{ranktwo}0.459 & 0.603 & 1.157 & 5.5 \\
    \hyperlap  & 22.4 & 54.6 & 85.9 & 1.012 & \cellcolor{rankone}0.217 & 1.129 & 6.0 \\
    \hyperdk   & 24.6 & 54.5 & 82.9 & 0.726 & 0.511 & 0.978 & 5.2 \\
    \thera     & 21.0 & 54.3 & 80.6 & \cellcolor{rankthree}0.704 & 0.484 & \cellcolor{rankthree}0.964 & \cellcolor{rankthree}4.3 \\
    \hycosbm   & \cellcolor{rankone}5.0  & \cellcolor{rankone}52.2 & 102.5 & \cellcolor{rankone}0.281 & \cellcolor{rankthree}0.403 & \cellcolor{ranktwo}0.932 & \cellcolor{rankone}3.0 \\
    \hyrec     & 24.7 & 55.7 & \cellcolor{rankthree}77.6 & 0.797 & 0.523 & 0.968 & 5.7 \\
    \midrule
    \textbf{\method}    & \cellcolor{ranktwo}9.3 & \cellcolor{ranktwo}53.0 & \cellcolor{rankone}66.2 & 0.967 & 1.670 & \cellcolor{rankone}0.524 & \cellcolor{ranktwo}3.7 \\
    \textbf{\method-CF} & 21.6 & 66.5 & 86.3 & 1.232 & 3.129 & 1.539 & 8.8 \\
    \bottomrule
\end{tabularx}
}
\vspace{0.1mm}
\subfloat[Yelp Bar (\method ranks \textbf{first} overall)]{
\begin{tabularx}{\columnwidth}{p{1.9cm}|Y{0.7cm}Y{0.7cm}Y{0.7cm}Y{0.7cm}Y{0.7cm}Y{0.7cm}|Y{0.5cm}}
    \toprule
    & \textbf{T2} & \textbf{T3} & \textbf{T4} & \textbf{HE} & \textbf{HOHE} & \textbf{NHS} & \textbf{A.R.} \\
    \midrule
    \hypercl   & 57.1 & 112.1 & 133.2 & 1.392 & 1.064 & 1.343 & 7.0 \\
    \hyperpa   & 54.7 & 114.4 & 139.9 & 1.090 & 1.391 & 1.283 & 6.8 \\
    \hyperff   & 54.6 & 115.2 & \cellcolor{rankthree}129.4 & \cellcolor{rankthree}0.635 & \cellcolor{rankone}0.700 & 1.430 & 4.8 \\
    \hyperlap  & \cellcolor{rankthree}54.1 & \cellcolor{rankthree}107.2 & 137.0 & 1.345 & \cellcolor{rankthree}0.978 & 1.325 & 5.2 \\
    \hyperdk   & 57.1 & 108.7 & 134.2 & 1.572 & 1.102 & 1.230 & 6.0 \\
    \thera     & 55.5 & \cellcolor{ranktwo}106.0 & 133.3 & 1.124 & 1.061 & \cellcolor{rankthree}1.221 & \cellcolor{ranktwo}4.3 \\
    \hycosbm   & \cellcolor{rankone}26.3 & 128.9 & 161.3 & \cellcolor{rankone}0.526 & 1.053 & \cellcolor{ranktwo}1.140 & \cellcolor{rankthree}4.7 \\
    \hyrec     & 58.3 & 111.3 & \cellcolor{rankone}123.8 & 1.343 & \cellcolor{ranktwo}0.953 & 1.225 & 5.0 \\
    \midrule
    \textbf{\method}    & \cellcolor{ranktwo}44.7 & \cellcolor{rankone}94.0 & 145.2 & \cellcolor{ranktwo}0.589 & 1.598 & \cellcolor{rankone}0.468 & \cellcolor{rankone}4.0 \\
    \textbf{\method-CF} & 61.7 & 113.9 & \cellcolor{ranktwo}127.9 & 0.827 & 2.955 & 1.630 & 7.2 \\
    \bottomrule
\end{tabularx}
}
\hfill
\subfloat[Devops (\method ranks \textbf{fourth} overall)]{
\begin{tabularx}{\columnwidth}{p{1.9cm}|Y{0.7cm}Y{0.7cm}Y{0.7cm}Y{0.7cm}Y{0.7cm}Y{0.7cm}|Y{0.5cm}}
    \toprule
    & \textbf{T2} & \textbf{T3} & \textbf{T4} & \textbf{HE} & \textbf{HOHE} & \textbf{NHS} & \textbf{A.R.} \\
    \midrule
    \hypercl & 1,257 & 3,450 & \cellcolor{rankthree}5,793 & \cellcolor{rankthree}4.89 & 30.86 & \cellcolor{ranktwo}8.45 & \cellcolor{rankthree}3.3 \\
    \hyperpa & 2,428 & 6,670 & 9,427 & 26.22 & 66.76 & 20.3 & 9.0 \\
    \hyperff & 2,382 & 6,605 & 9,204 & 25.71 & 66.74 & 20.89 & 8.0 \\
    \hyperlap & \cellcolor{rankthree}1,246 &\cellcolor{rankthree}3,363 & \cellcolor{ranktwo}5,780 & \cellcolor{ranktwo}4.82 & \cellcolor{rankthree}30.63 & \cellcolor{rankthree}8.99 & \cellcolor{ranktwo}2.7 \\
    \hyperdk & \cellcolor{rankone}1,136 & \cellcolor{ranktwo}2,765 & \cellcolor{rankone}4,787 & 6.86 & \cellcolor{ranktwo}28.56 & 10.27 & \cellcolor{rankone}2.3 \\
    \thera & 2,387 & 6,567 & 9,136 & 25.9 & 65.53 & 20.52 & 7.5 \\
    \hycosbm & \cellcolor{ranktwo}1,226 & \cellcolor{rankone}2,696 & 7,027 & \cellcolor{rankone}4.52 & 53.83 & 25.43 & 3.8 \\
    \hyrec & 2,394 & 6,736 & 8,943 & 23.31 & 59.88 & 15.96 & 7.2 \\
    \midrule
    \textbf{\method} & 1,714 & 5,182 & 8,038 & 12.81 & \cellcolor{rankone}19.75 & \cellcolor{rankone}6.59 & 3.7 \\
    \textbf{\method-CF} & 2,158 & 6,330 & 9,199 & 30.05 & 77.31 & 14.22 & 7.5 \\
    \bottomrule
\end{tabularx}
}
\vspace{0.1mm}
\subfloat[Patents (\method ranks \textbf{fifth} overall)]{
\begin{tabularx}{\columnwidth}{p{1.9cm}|Y{0.7cm}Y{0.7cm}Y{0.7cm}Y{0.7cm}Y{0.7cm}Y{0.7cm}|Y{0.5cm}}
    \toprule
    & \textbf{T2} & \textbf{T3} & \textbf{T4} & \textbf{HE} & \textbf{HOHE} & \textbf{NHS} & \textbf{A.R.} \\
    \midrule
    \hypercl & 6,586 & \cellcolor{rankone}13,091 & \cellcolor{rankone}16,515 & 64.30 & \cellcolor{rankthree}105.59 & \cellcolor{ranktwo}103.33 & \cellcolor{rankone}2.5 \\
    \hyperpa & 13,538 & 27,184 & 31,883 & 268.45 & 466.07 & 233.28 & 9.0 \\
    \hyperff & 12,069 & 26,032 & 29,718 & 262.36 & 440.24 & 230.16 & 6.2 \\
    \hyperlap & \cellcolor{rankthree}6,551 & \cellcolor{ranktwo}13,417 & \cellcolor{ranktwo}16,663 & \cellcolor{rankthree}61.98 & \cellcolor{ranktwo}104.40 & \cellcolor{rankthree}104.40 & \cellcolor{rankone}2.5 \\
    \hyperdk & \cellcolor{ranktwo}6,262 & 15,041 & 23,320 & \cellcolor{ranktwo}53.40 & \cellcolor{rankone}85.05 & \cellcolor{rankone}91.81 & \cellcolor{rankone}2.5 \\
    \thera & 12,856 & 26,900 & 30,711 & 266.45 & 464.98 & 232.26 & 7.8 \\
    \hycosbm & \cellcolor{rankone}5,828 & \cellcolor{rankthree}13,984 & \cellcolor{rankthree}18,112 & \cellcolor{rankone}48.52 & 265.48 & 230.81 & 3.3 \\
    \hyrec & 14,769 & 26,521 & 29,705 & 263.42 & 442.54 & 228.55 & 7.0 \\
    \midrule
    \textbf{\method} & 7,292 & 17,557 & 22,532 & 65.32 & 109.78 & 134.08 & 4.5\\
    \textbf{\method-CF} & 14,298 & 27,920 & 31,497 & 268.81 & 468.97 & 236.31 & 9.7 \\
    \bottomrule
\end{tabularx}
}
\hfill
\subfloat[Average Rank over Nine Datasets (\method ranks \textbf{first} overall)]{
\begin{tabularx}{\columnwidth}{p{1.9cm}|Y{0.7cm}Y{0.7cm}Y{0.7cm}Y{0.7cm}Y{0.7cm}Y{0.7cm}|Y{0.5cm}}
    \toprule
    & \textbf{T2} & \textbf{T3} & \textbf{T4} & \textbf{HE} & \textbf{HOHE} & \textbf{NHS} & \textbf{A.R.} \\
    \midrule
    \hypercl   & 6.9 & 5.7 & 5.0 & 6.6 & 5.3 & 6.6 & 6.2 \\
    \hyperpa   & 7.8 & 8.6 & 7.4 & 7.7 & 8.0 & 7.2 & 9.7 \\
    \hyperff   & 5.2 & 6.3 & 6.2 & 5.1 & 5.4 & 6.1 & 5.7 \\
    \hyperlap  & 5.3 & 5.3 & 5.3 & 6.2 &\cellcolor{rankone}2.8 & 5.6 & 4.7 \\
    \hyperdk   & 6.1 &\cellcolor{ranktwo}3.9 &\cellcolor{rankthree}4.7 &\cellcolor{rankthree}4.3 & 5.4 &\cellcolor{rankthree}4.6 &\cellcolor{ranktwo}3.8 \\
    \thera     &\cellcolor{rankthree}4.9 &\cellcolor{rankthree}4.0 & 5.8 & 5.2 &\cellcolor{ranktwo}4.0 & 5.0 &\cellcolor{ranktwo}3.8 \\
    \hycosbm   &\cellcolor{rankone}1.2 & 4.8 & 6.4 &\cellcolor{ranktwo}4.1 & 6.0 & 6.7 & 5.3 \\
    \hyrec     & 7.7 & 7.1 &\cellcolor{rankone}3.9 & 5.6 & 5.1 &\cellcolor{ranktwo}4.0 & 5.0 \\
    \midrule
    \textbf{\method}    &\cellcolor{ranktwo}2.9 &\cellcolor{rankone}2.1 &\cellcolor{rankone}3.9 &\cellcolor{rankone}3.4 &\cellcolor{rankthree}4.4 &\cellcolor{rankone}1.4 &\cellcolor{rankone}1.5 \\
    \textbf{\method-CF} & 7.0 & 7.2 & 6.3 & 6.8 & 8.4 & 7.9 & 9.0 \\
    \bottomrule
\end{tabularx}
}
\end{table*}

\smallsection{Baselines.} 
We consider eight baseline \red{generative models}: \hypercl~\cite{lee2021hyperedges}, \hyperpa~\cite{do2020structural}, \hyperff~\cite{kook2020evolution}, \hyperlap~\cite{lee2021hyperedges}, \hyperdk~\cite{nakajima2021randomizing}, \thera~\cite{kim2023transitive}, \hycosbm~\cite{badalyan2024structure}, \hyrec~\cite{choe2025kronecker}. 
Among the baselines, \hycosbm explicitly utilizes node attributes.
For \hypercl, \hyperlap, and \hyperdk,
since node identities are preserved in the generated hypergraphs, we assign node attributes based on their correspondence to the original nodes.
For \hyperpa, \hyperff, \thera, and \hyrec, where node identities are not preserved during the generation process, we assign node attributes randomly. 
The hyperparameter search spaces for \red{our method} and the baselines are detailed in Appendix X~\cite{online}. 

\smallsection{Evaluation.}
We compare \method and the baselines in terms of their ability to reproduce the structure-attribute interplay observed in real-world hypergraphs based on the metrics described in Section~\ref{sec:prelim:measures}. 
For type-$s$ affinity ratio scores, we evaluate hypergraph generators by comparing the sum of log scaled differences between the ground truth and the generated hypergraphs over all $t \in [1, s]$ and all \red{attributes.}
For (higher-order) hyperedge entropy and node homophily score, since they are presented as distributions for each attribute, we calculated the sum of Wasserstein Distance 
between the ground-truth and generated hypergraphs. 
We denote type-$s$ affinity ratio scores as $\textbf{Ts}$, hyperedge entropy as $\textbf{HE}$, higher-order hyperedge entropy as $\textbf{HOHE}$, and node homophily score as $\textbf{NHS}$.
\red{For each metric and dataset, we compute both the raw values and the rankings of all compared models.}

\smallsection{Machines.}
We conducted all experiments on a server with RTX A6000 GPUs.

\subsection{Performance Comparison}
\label{sec:exp:perf}

As shown in Table~\ref{tab:interplayeval}, \method achieves the best average rank in 5 out of 9 datasets.
Averaged over nine datasets, \method ranks first in four metrics (type-3 affinity score, type-4 affinity score, hyperedge entropy, and node homophily score), second in type-2 affinity score, and third in higher-order hyperedge entropy.
These results demonstrate the overall superiority of \method in capturing the structure–attribute interplay of real-world hypergraphs.

However, there are cases where \method underperforms.
First, its relatively low performance in the online Q\&A domain datasets (Devops and Patents) \red{is likely} due to the weak correlation between attribute and structure in these datasets.
In the datasets, baselines that \red{explicitly} preserve degree and size distributions, including \hypercl, \hyperlap, and \hyperdk, perform well.
Second, \hyperlap and \thera tend to preserve higher-order hyperedge entropy (HOHE) more effectively than \method. This is because, due to the label-propagation steps, HOHE is strongly influenced by hyperedge overlaps, which \hyperlap and \thera explicitly target.


\begin{figure}[t] 
    \vspace{-3mm}
    \centering
    \includegraphics[width=0.8\columnwidth]{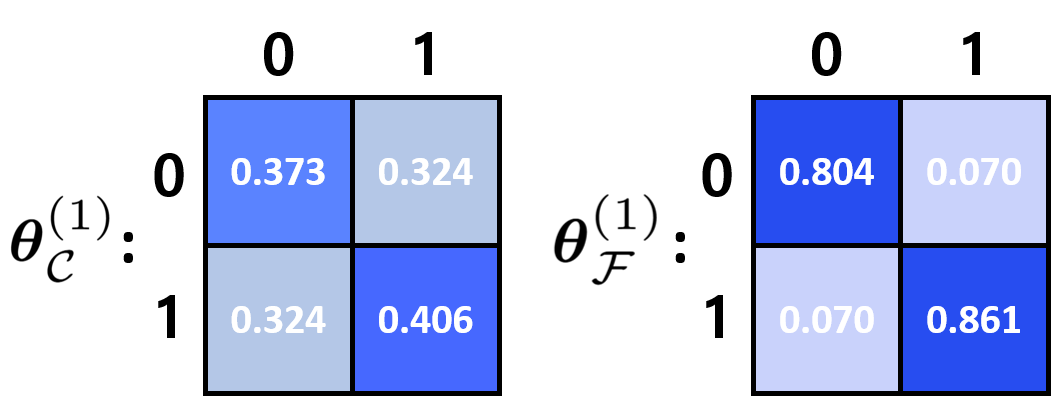}
    \caption{$\mat{\theta}_{\mathcal{C}}^{(1)}$ and $\mat{\theta}_{\mathcal{F}}^{(1)}$  estimated by \fitwattr
    on the Amazon Music dataset. \fitwattr captures homophily by assigning higher affinities to same attribute value pairs ($0 \leftrightarrow 0$ and $1 \leftrightarrow 1$) than to different attribute value pairs ($0 \leftrightarrow 1$).}
    \label{fig:affinity_matrix}
\end{figure}

\subsection{Case Study}
\label{sec:exp:case}
To gain deeper insight into the effectiveness of \method, we conduct a case study on the Amazon Music dataset~\cite{ni2019justifying}. 
In the Amazon Music dataset, a value of the \red{first node attribute} indicates whether a reviewer has reviewed music in the New York Blues genre \red{(1) or not (0).}
For this attribute, we compare the structure-attribute interplay metrics between (1) the ground truth hypergraph, (2) the hypergraph generated by \hypercl, and (3) the hypergraph generated by \method with \fitwattr.
As shown in Figure~\ref{fig:intro_example} in Section~\ref{sec:intro}, the distributions of hyperedge entropy and node homophily scores are highly skewed toward 0 and 1, respectively.
Additionally, \red{the type-(4, 4) affinity ratio score is high
for both attribute values 0 and 1}, indicating that many size 4 hyperedges consist of nodes sharing the same value \red{for the first attribute.}
These results suggest that nodes in the Amazon Music dataset exhibit strong homophily with respect to the first attribute, and that the node attribute plays a crucial role in hypergraph formation.
Whereas the baseline model,  \hypercl, fails to capture this structure–attribute interplay, \method successfully captures it through the use of affinity matrices shown in Figure~\ref{fig:affinity_matrix}.

\begin{figure*}[t]
\vspace{-10mm}
    \centering
    \subfloat[Cora\label{fig:cora_topology}]{
        \includegraphics[width=0.49\textwidth]{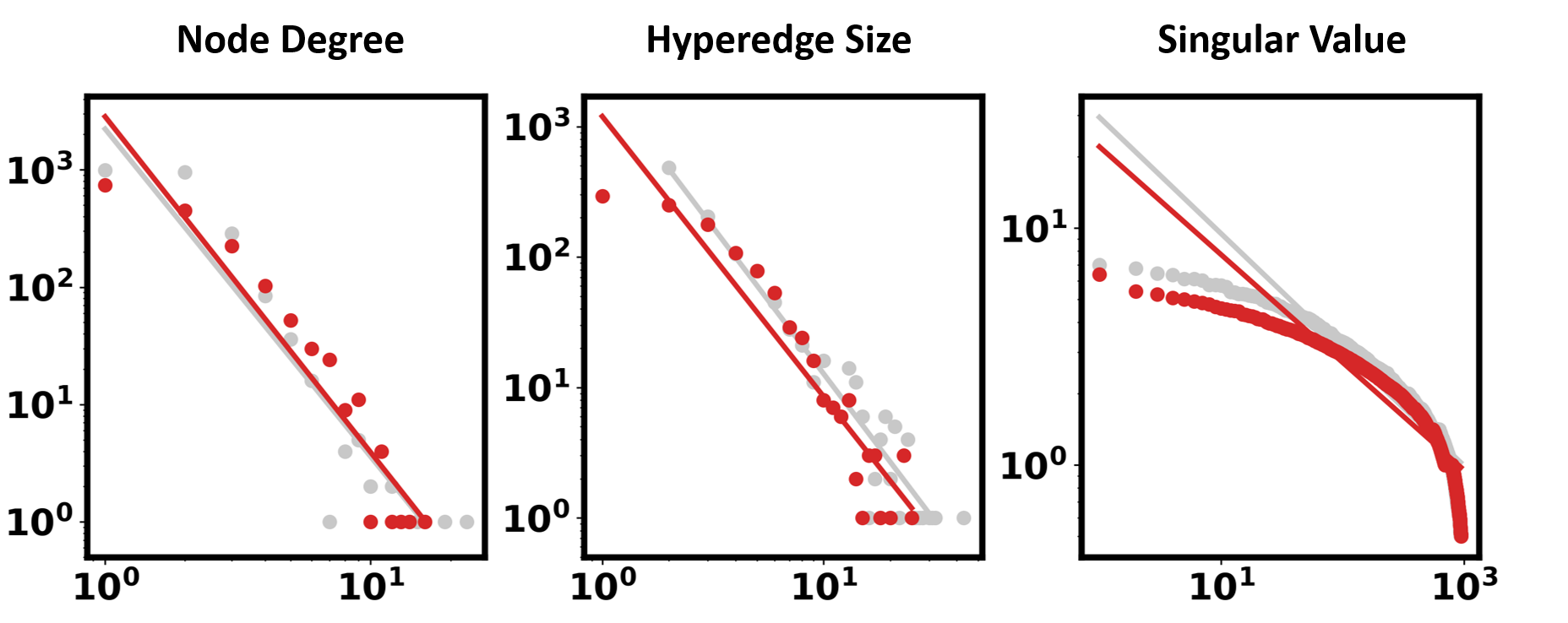}
        \vspace{-0.5mm}
    }
    \subfloat[Workspace\label{fig:workspace_topology}]{
        \includegraphics[width=0.49\textwidth]{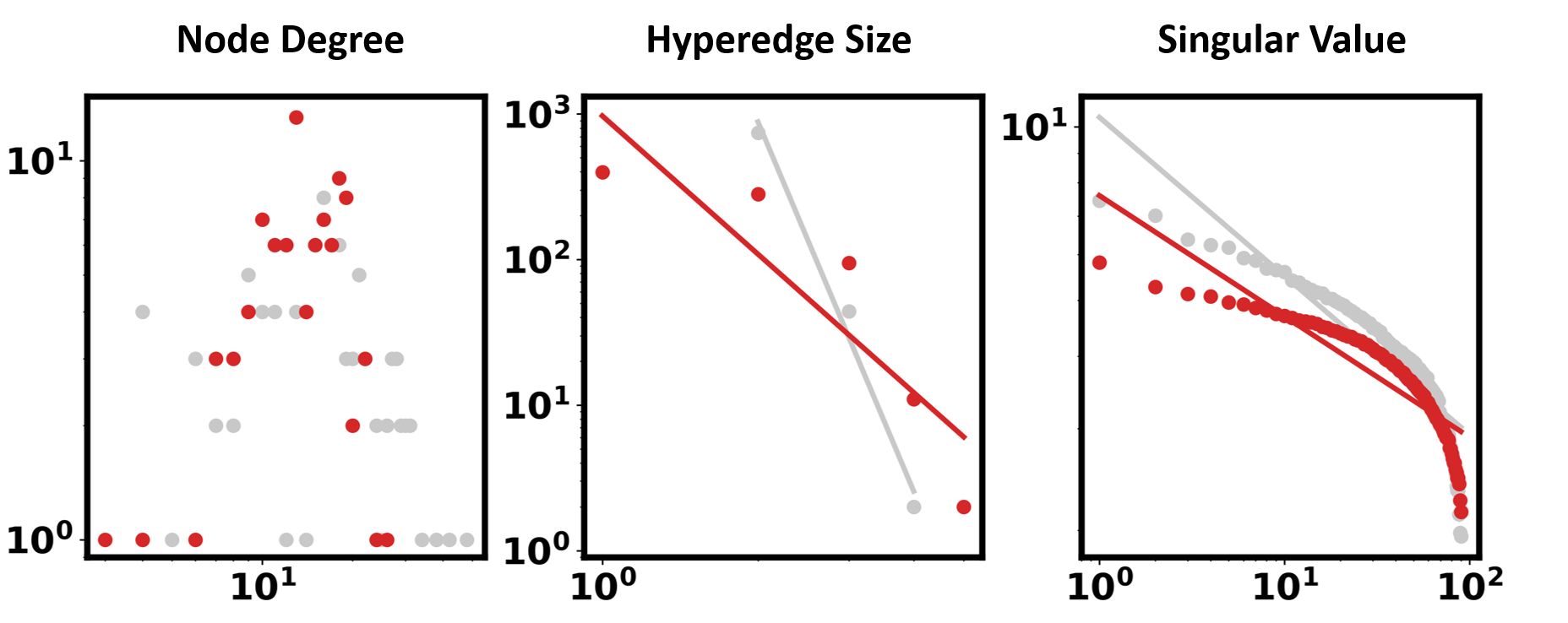}
        \vspace{-0.5mm}
    }
    \vspace{-4mm}
    \subfloat[Yelp Restaurant\label{fig:restaurant_topology}]{
        \includegraphics[width=0.49\textwidth]{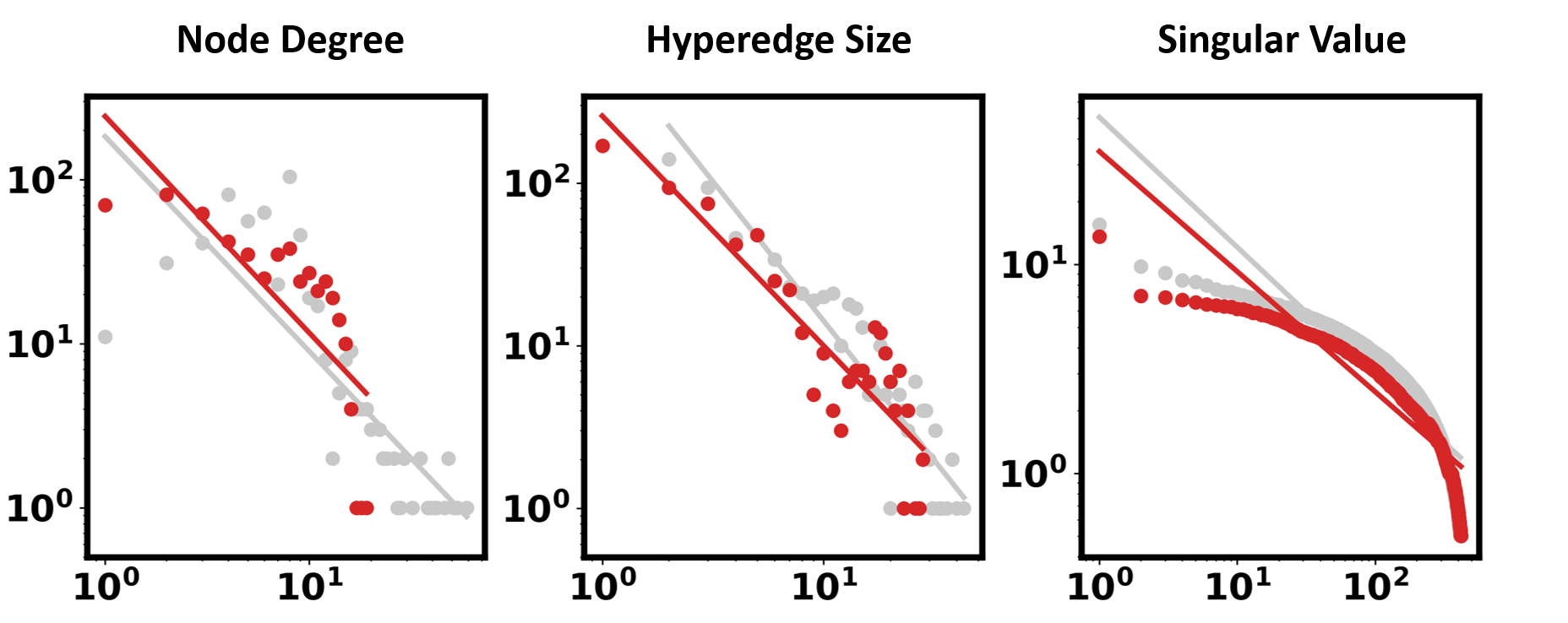}
        \vspace{-0.5mm}
    }
    \subfloat[Devops\label{fig:devops_topology}]{
        \includegraphics[width=0.49\textwidth]{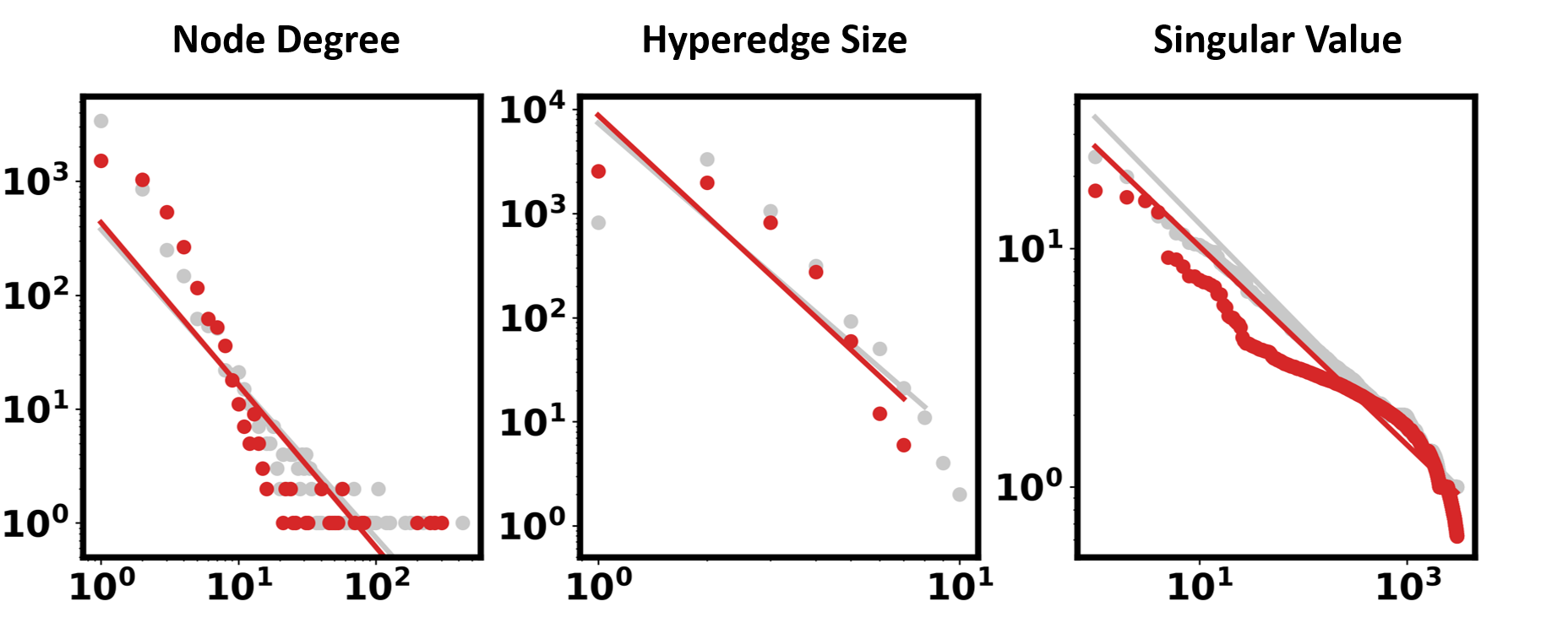}
        \vspace{-0.5mm}
    }
    \vspace{1mm}
    \caption{Structural patterns of real-world hypergraphs (\textcolor{real}{gray}) and those generated by \method tuned by \fitwattr (\textcolor{method}{red}). \red{These results demonstrate that, despite its focus on  structure–attribute interplay, \method also successfully reproduces purely structural patterns.}}
    \label{fig:topology}
\end{figure*}

\subsection{Ablation Study}
\label{sec:exp:ablation}
To assess the contribution of the core–fringe node hierarchy in \method, we compare its performance with a variant, \method-CF, which omits this hierarchical structure. 
In \method-CF, each hyperedge is generated by sampling a seed node from the entire node set and attaching additional nodes directly, without distinguishing between core and fringe roles (and therefore without collective consideration of multiple core nodes).
As reported in Table~\ref{tab:interplayeval}, \method consistently outperforms \method-CF across all datasets, showing the importance of the core–fringe hierarchy in realistic hyperedge formation.

\subsection{Scalability Analysis}
\label{sec:exp:scale}
\begin{figure}[t]
    \vspace{-5.5mm}
    \centering
    \captionsetup[subfloat]{justification=centering}
    \subfloat[\fitwattr (fitting) w.r.t. number of hyperedge\label{fig:scale_edge_fit}]{
        \includegraphics[width=0.22\textwidth]{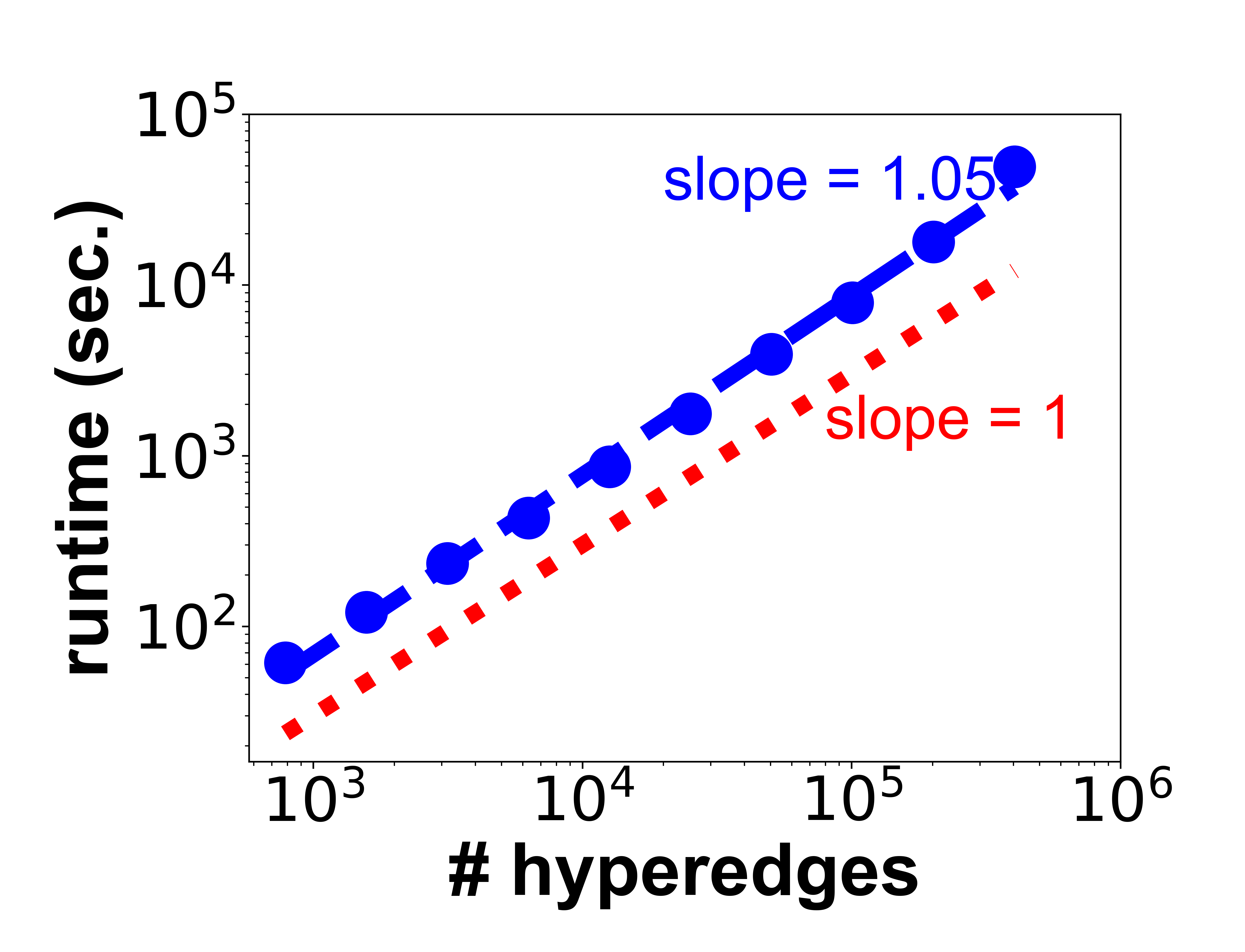}
    }
    \subfloat[\method (generation) w.r.t. number of hyperedge\label{fig:scale_edge_gen}]{
        \includegraphics[width=0.22\textwidth]{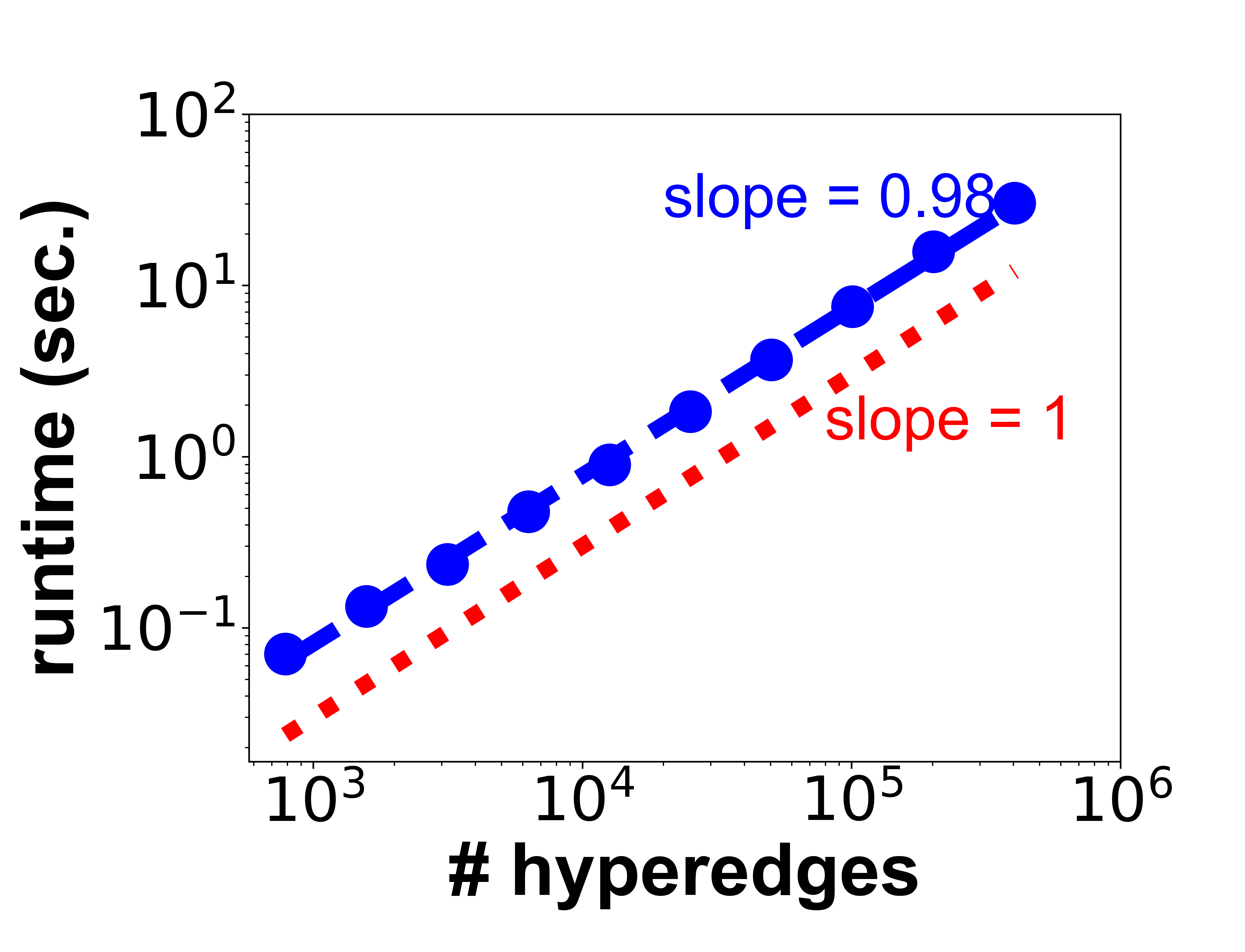}
    }
    \vspace{-4mm}
    \subfloat[\fitwattr (fitting) w.r.t. number of attributes\label{fig:scale_attr_fit}]{
        \includegraphics[width=0.22\textwidth]{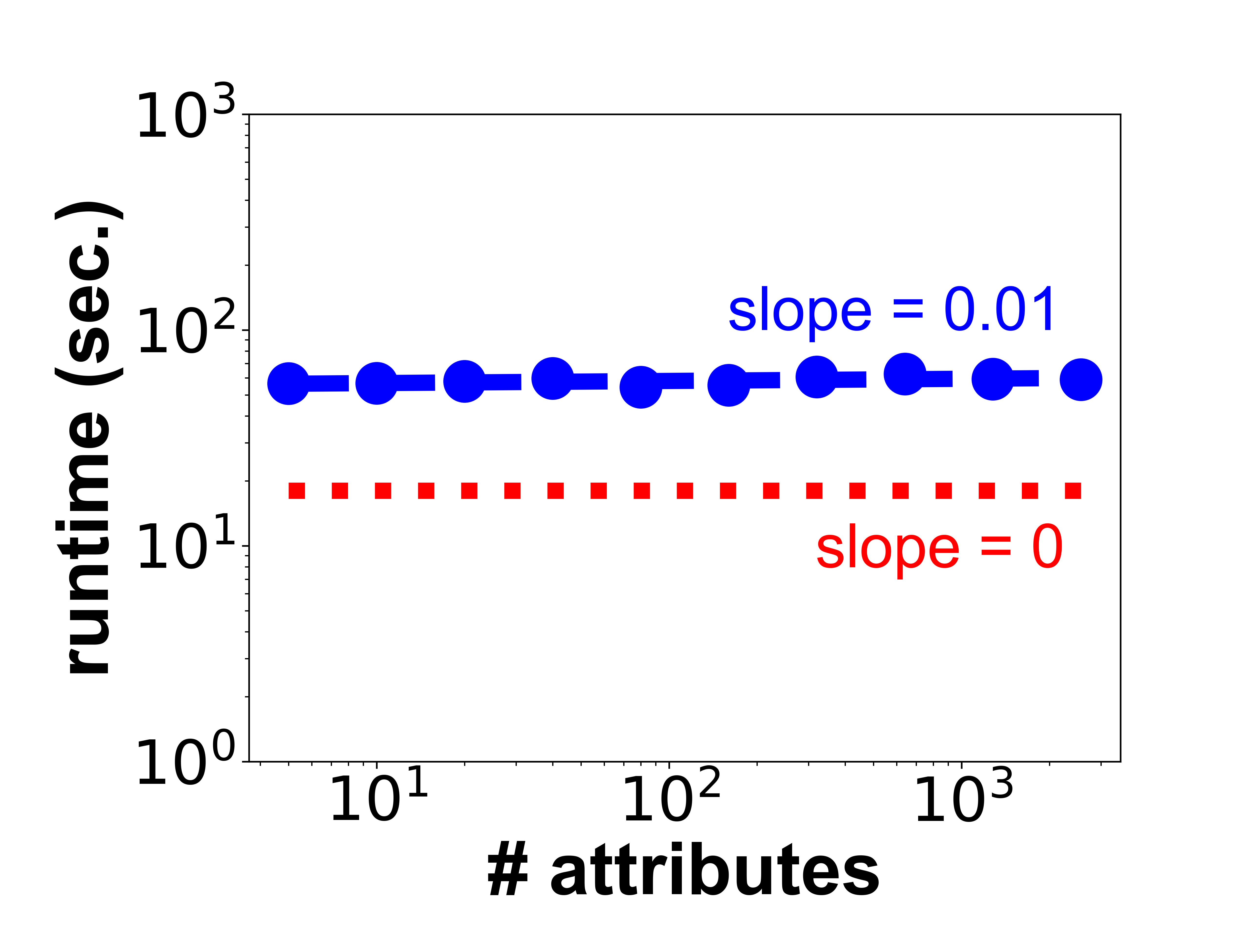}
    }
    \subfloat[\method (generation) w.r.t. number of attributes \label{fig:scale_attr_gen}]{
        \includegraphics[width=0.22\textwidth]{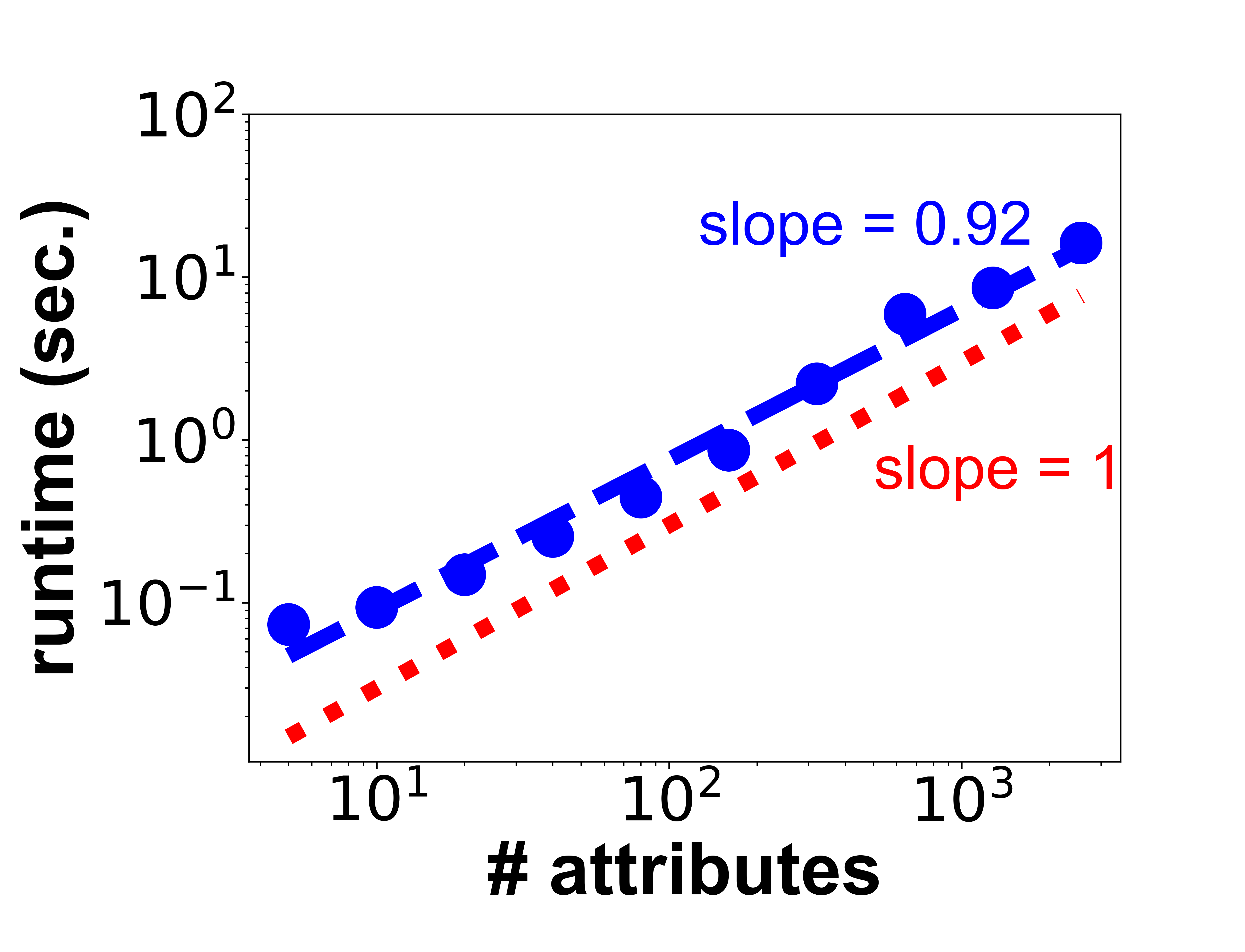}
    }
    \vspace{0mm}
    \caption{\fitwattr scales nearly linearly with the number of hyperedges and \red{remains almost constant} with respect to the number of attributes. \method scales nearly linearly with both the number of hyperedges and attributes.}
    \label{fig:scalabilty}
\end{figure}

 We evaluate the scalability of our fitting algorithm, \fitwattr, and our generative model, \method, with respect to the number of hyperedges and  attributes.
To this end, we scale up the Contact-Workspace dataset by factors ranging from 2 to 512.
Figure~\ref{fig:scalabilty} shows the runtime of fitting and generation as functions of the number of hyperedges and attributes, plotted on a log–log scale. 
The proposed fitting algorithm, \fitwattr, scales nearly linearly with the number of hyperedges and remains almost constant with respect to the number of attributes. 
The proposed generative model, \method, scales near-linearly with both the number of hyperedges and the number of attributes.
These results demonstrate the scalability of both the fitting and generation components of our framework.

\subsection{Further Analysis} \label{sec:result:further}
Additionally, we examine the structural patterns of hypergraphs generated by \method in terms of (1) node degrees, (2) hyperedge sizes, and (3) singular values. 
These properties are known to follow heavy-tailed distributions in real-world hypergraphs~\cite{ko2022growth}.
In Figure~\ref{fig:topology}, we visually demonstrate that \method with \fitwattr reproduces such structural patterns, \red{which closely resemble those of the input hypergraphs.}
We also present a detailed evaluation on structural metrics in Appendix XI~\cite{online}, \red{where we show that, despite its focus on structure–attribute interplay, \method achieves competitive results on purely structural patterns (with an average rank of 4.6, when compared with 8 structure-focused methods).}

%% file: 070conclusion.tex
In this work, we proposed \method, a stochastic generative model for attributed hypergraphs that reproduces \red{realistic interplay between structure and node attributes.} 
By leveraging a two-level node hierarchy, core and fringe nodes, \method formulates hyperedge generation as sequential attachment of nodes \red{(first cores, then fringes)}, where attachment probabilities are governed by node attributes. 
We also introduced \fitwattr, a parameter estimation algorithm that fits \method to \red{a given hypergraph} by estimating affinity matrices and seed core probabilities.
Through extensive experiments on nine real-world hypergraphs across four diverse domains, we demonstrated that \method with \fitwattr more accurately \red{reproduces} the structure–attribute interplay than eight existing hypergraph generative models across six metrics. 

\smallsection{Relevance to Data Mining and Broad Impact.}
\red{The proposed framework aligns with the fundamental goal of data mining: finding models that explain complex, large-scale data. By generating realistic data, it supports diverse applications in domains where hypergraphs naturally arise, enabling statistical analysis, simulation, and data anonymization.}

\smallsection{Future Directions.}
{
We plan to extend our model and fitting algorithm to handle more complex attributes, including continuous ones. We also aim to further enhance their scalability to support even larger, web-scale hypergraphs.
}
